RESEARCH

# A `Downstream` and Vertexing Algorithm for Long Lived Particles (LLP) Selection at the First High Level Trigger (HLT1) of LHCb

V. Kholoimov[1] · B. Kishor Jashal[1,2] · A. Oyanguren[1] · V. Svintozelskyi[1] · J. Zhuo[1]

## Abstract
A new algorithm has been developed at LHCb which is able to reconstruct and select very displaced vertices in real time at the first level of the trigger (HLT1). It makes use of the Upstream Tracker (UT) and the Scintillator Fiber detector (SciFi) of LHCb and it is executed on GPUs inside the Allen framework. In addition to an optimized strategy, it utilizes a Neural Network (NN) implementation to increase the track efficiency and reduce the ghost rates, with very high throughput and limited time budget. Besides serving to reconstruct $K^0_S$ and $\Lambda$ particles from the Standard Model, the `Downstream` algorithm and the associated two-track vertexing could largely increase the LHCb physics potential for detecting long-lived particles during the Run 3.

**Keywords** LHCb · HLT1 · GPUs · Downstream · LLPs

## Introduction

The LHCb forward spectrometer is one of the main detectors at the Large Hadron Collider (LHC) accelerator at CERN, with the primary purpose of searching for new physics through studies of CP-violation and heavy-flavour hadron decays. It has been operating during its Run 1 (2011–2012) and Run 2 (2015–2018) periods with very high performance, recording an integrated luminosity of 9 $\text{fb}^{-1}$ at center-of-mass energies of 7, 8, and 13 TeV and delivering a plethora of accurate physics results and new particles discoveries. One of the main issues concerning the present Run 3 was that, even if many physics results are statistically limited, a direct increase of the luminosity provided by the accelerator was not directly translated into an increase of the selected events of interest, and a new full-software-based selection (trigger) strategy was necessary [1], named in the following High Level Trigger (HLT). This new trigger paradigm has been very successful allowing LHCb to collect 10 $\text{fb}^{-1}$ at 13 TeV during 2023–2024. Due to the computing timing constraints, during previous runs the first stage of the trigger was based on fast partial particle tracking reconstruction, since the large hit occupancy made the throughput decrease by two orders of magnitude. At that time Intel Xeon E5-2630 v3 CPUs were used for performing the reconstruction and selection tasks. A new algorithm called `Downstream` has been developed in this work, which makes use of the new trigger scheme based on A5000 NVIDIA GPUs.

✉ J. Zhuo
jiahui.zhuo@cern.ch

1 Instituto de Física Corpuscular (IFIC), University of Valencia-CSIC, Valencia, Spain

2 Rutherford Appleton Laboratory (RAL), Oxford, UK

### The LHCb Detector

The upgraded LHCb detector, operational at present during the Run 3 of the LHC, has implied a major change in the experiment. The detectors have been almost completely renewed to allow running at an instantaneous luminosity five times larger than that of the previous running periods, in particular using new readout architectures. A full software trigger executed on Graphic Processor Units (GPU) also represents one of the main features of the new LHCb design, allowing the reconstruction and selection of events in real time and widening the physics reach of the experiment. The main characteristics of the new LHCb detector are detailed in Ref. [2], and summarised in the following. As compared to the previous detector [3], one of the most important improvements concerns the new tracking system. The LHCb is comprised of a three subdetector tracking system, a particle identification system, based on two-ring imaging Cherenkov detectors, hadronic and electromagnetic calorimeters, and four muon chambers.

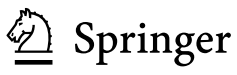

The tracking system of the LHCb experiment consists of three subsystems, VELO, UT, and SciFi, which are responsible for reconstructing charged particles. A dipole magnet, with a bending power of 4 Tm, is also necessary to curve particle trajectories to measure their momentum, $p$. The magnetic field vector $\vec{B}$ is oriented vertically, so tracks are bent mostly in the horizontal plane. The magnet polarity can be inverted, and it is used to control systematic effects coming from detector and alignment imperfections.

The VELO is based on pixelated silicon sensors and is critical for determining the decay vertices of $b$ and $c$ flavored hadrons. The UT contains vertically segmented silicon strips and continues the tracking upstream of the VELO. It is also used to determine the momentum of charged particles and is useful to remove low-momentum tracks from being extrapolated downstream, thus speeding up the software trigger by about a factor of three. Tracking after the magnet is handled by the new scintillating fiber-based detector SciFi. Two Ring Imaging Cherenkov (RICH) detectors supply particle identification. RICH1 detects mainly low-momentum particles, from 2 to 10 GeV/c, and RICH2 is sensitive to higher momentum ones, up to 100 GeV/c. The Electromagnetic Calorimeter (ECAL) identifies electrons and reconstructs photons such as from neutral pions. The Hadronic Calorimeter (HCAL) measures the energy deposition of hadrons, while the four muon chambers M2–M5 are mostly used for muon identification. The angular coverage of the LHCb detectors ranges from $2 < \eta < 5\,(0.77° < \theta < 15.4°)$. Figure 1 shows the LHCb upgrade detector.

Sketches of the main detectors involved in the `Downstream` algorithm are also shown in Figs. 2 and 3.

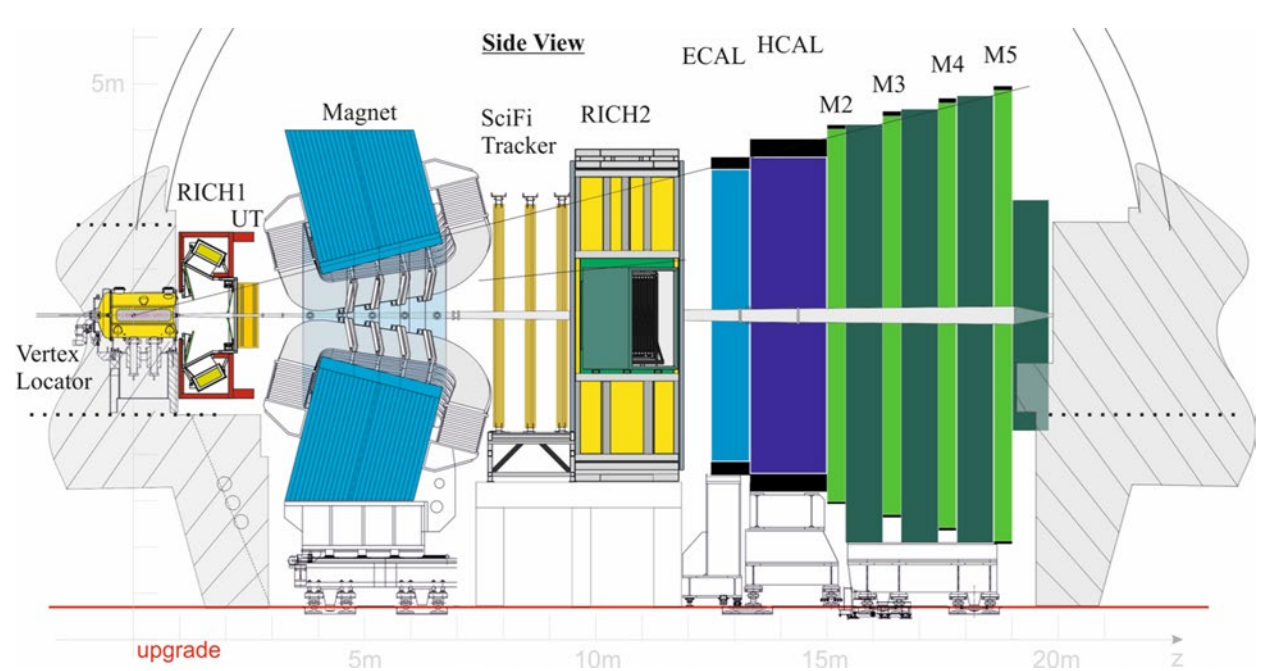

**Fig. 1** New LHCb detector operating during the Run 3 [2]. The LHCb coordinate system is a right-handed cartesian system with the $z$ axis aligned with the beamline and pointing downstream the detector, and the $y$ axis pointing upwards

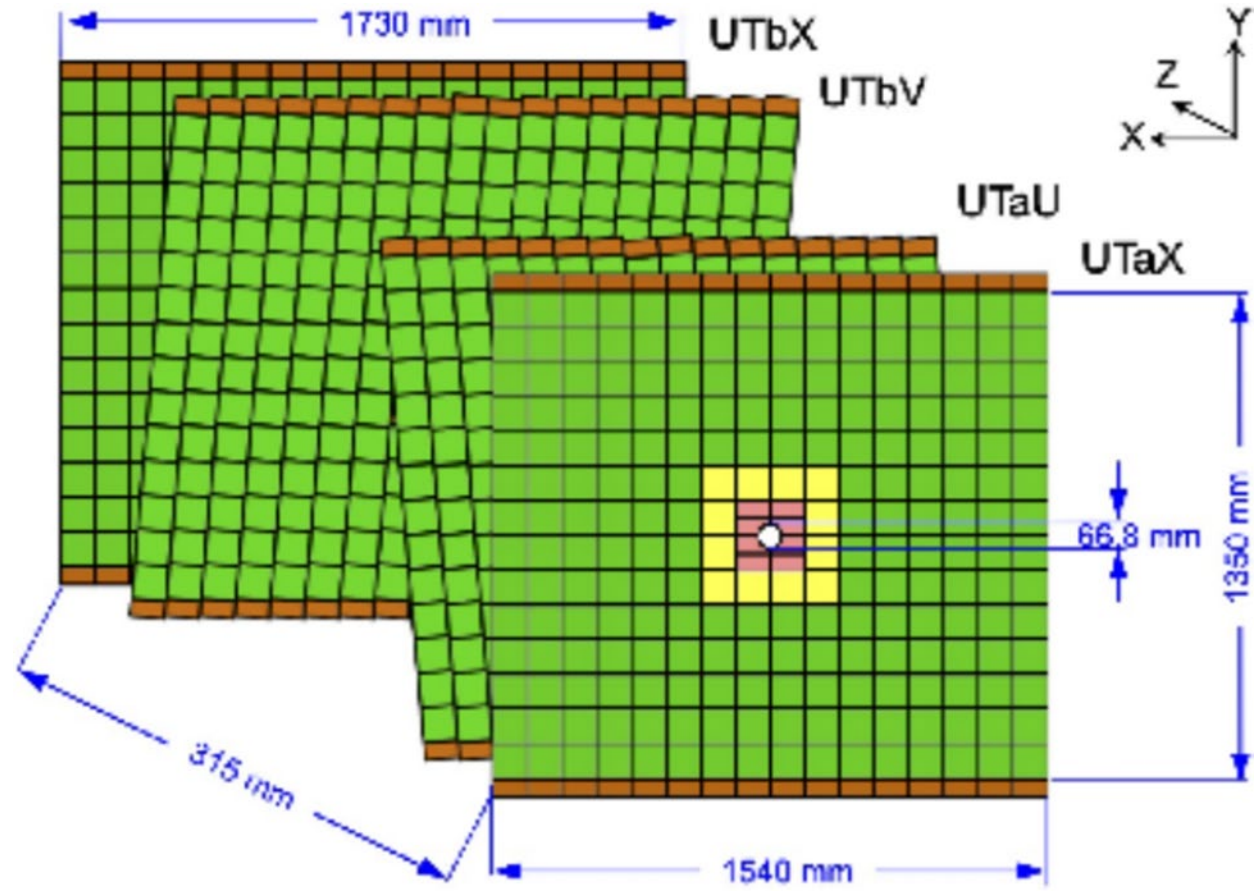

**Fig. 2** Sketch of the four silicon planes comprising the UT detector, with indicative dimensions. The first ("a") station is composed of an $x$-measuring layer (UTaX) with vertical strips and a stereo layer (UTaU) with strips inclined by 5°. The second ("b") station is similar, with first a stereo layer (UTbV) with opposite inclination, and a layer with vertical strips (UTbX) [2]

## Track Types at LHCb

Several track types are defined depending on the subdetectors involved in the reconstruction, as shown in Fig. 4.

The main track types considered for physics analyses are

*Long* tracks: they have information from at least the VELO and the SciFi, and possibly the UT. These are the main tracks used in physics analyses and at all stages of the trigger;

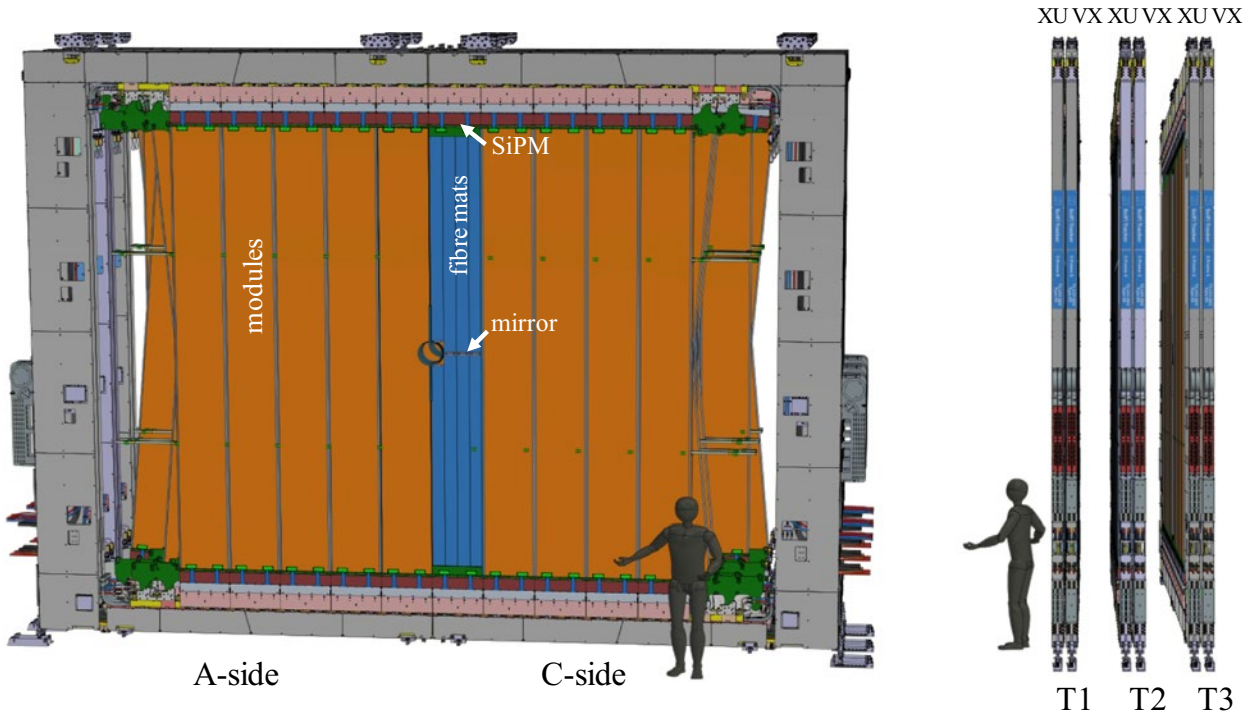

**Fig. 3** Front and side views of the SciFi detector. Twelve detection planes are arranged in three stations with 4 layers each in an $X{-}U{-}V{-}X$ configuration. The $X$ layers have their scintillator fibres oriented vertically and are used for determining the deflection of the charged particle tracks caused by the magnetic field. The inner two stereo layers, $U$ and $V$, have their fibres rotated by $\pm 5°$ in the plane of the layer for reconstructing the vertical position of the track hit [2]

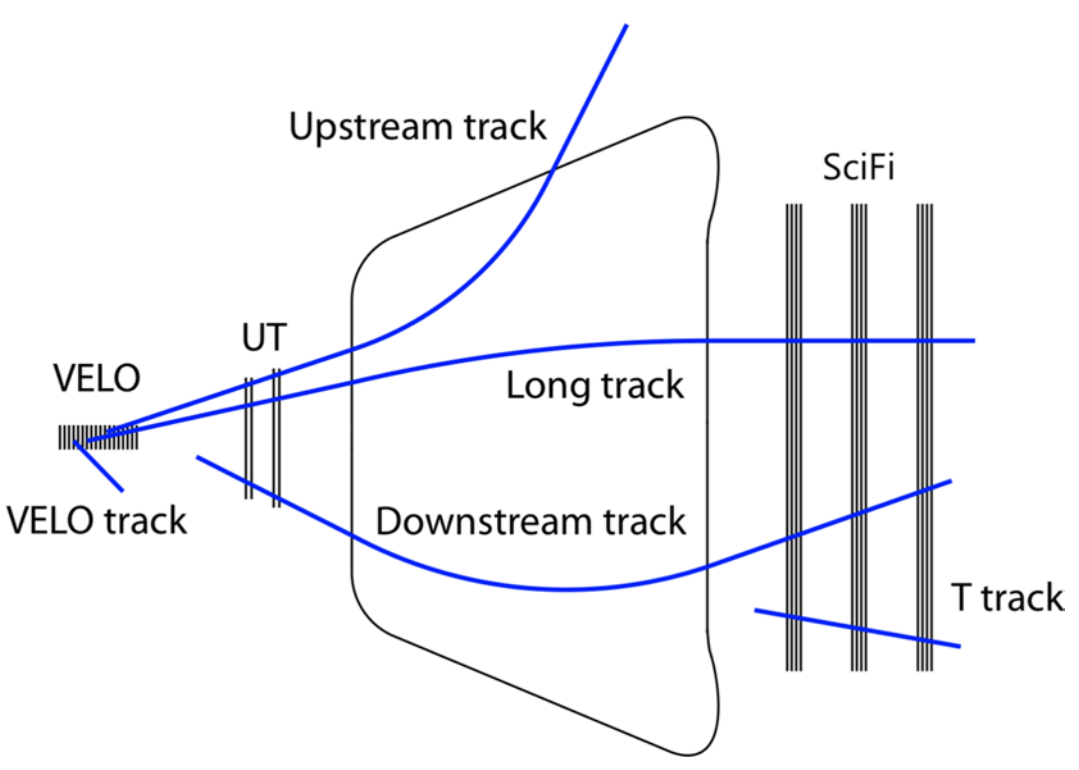

**Fig. 4** Definition of the particle track types in the LHCb experiment, according to which detectors are hit. The different tracker layers and the magnet in the center are sketched

*Downstream* tracks: they have information from the UT and the SciFi, but not VELO. They typically correspond to decay products of $K^0_S$ and $\Lambda$ hadron decays;
*T* tracks: they only have hits from the SciFi. They are typically not included in physics analysis. Nevertheless, their potential for physics has been recently probed [4, 5].

When simulating collision data using the Pythia [6], `Evt-Gen` [7] and Geant4 [8] packages, particle tracks meeting certain thresholds are defined to be *reconstructible* and have an assigned type according to the sub-detector reconstructibility. This is, in turn, based on the existence of reconstructed detector digits or clusters in the emulated detector, which are matched to simulated particles [9]. The reconstructibility in LHCb sub-detectors is defined as follows [9]:

- VELO reconstructible: at least 3 VELO sensors with at least 1 hit each.
- UT reconstructible: at least 2 hits, with 1 hit in layer one or two, and 1 hit in layer three or four. The hits can be axial or stereo.
- SciFi reconstructible: at least 1 axial and 1 stereo hit in each of the 3 SciFi stations.

Requirements for *long* tracks imply VELO and SciFi reconstructibility, *downstream* tracks must satisfy the UT and SciFi reconstructibility, and *T*-tracks only require the SciFi one.

### The HLT1 Trigger

The trigger system of the LHCb detector in Run 3 is software based and comprises two levels: HLT1 and HLT2, described in detail in Refs. [1, 10]. The HLT1 level has to be executed at a 30 MHz rate and, as such, suffers from heavy constraints on timing for event reconstruction. It performs partial event reconstruction to reduce the data rate. Tracking algorithms play a key role in fast event decisions, and the fact that they are inherently parallelisable processes suggests a way to increase trigger performance. Thus, the HLT1 has been implemented on a number of GPUs within the `Allen` software project [11], which allows to manage 40 Tbits/s raw data rate from the upgraded LHCb detector and reduces it by a factor of 30. After this initial selection, the data is passed to a buffer system, which enables "real-time" calibration and alignment of the detector, usually within a few minutes. Once completed, the calibration is applied to both HLT1 and HLT2 in data-taking. This output is used for the full and improved event reconstruction carried out by HLT2.

Due to timing constraints, the LHCb implementation in the HLT1 stage has been based on partial reconstruction and focuses solely on *long* tracks, i.e., tracks that have hits in the VELO. This trigger thus significantly affects the identification of particles with long lifetimes, particularly for LLP searches in LHCb, where some of the final-state particles are created further than roughly a metre away from the proton–proton interaction point and thus outside of the VELO acceptance. A new algorithm [12] has been developed and implemented to widen the reach of particle lifetimes of the HLT1 system, and is explained in the following.

## The `Downstream` Algorithm

`Downstream` is a fast and performant algorithm designed to reconstruct tracks which do not have hits in the VELO detector.

### Algorithm Design

The algorithm is based on the extrapolation of SciFi seeds (or *tracklets*) to the UT detector, including the effect of the magnetic field in the $x$ direction. Only seeds that have not been used for *long* track reconstruction are considered as *downstream* candidates. In Fig. 5, the flow chart of the algorithm is outlined.

SciFi *tracklets* are obtained using a standalone algorithm called `HybridSeeding` [13], modified for GPU execution in parallel. It performs a pattern recognition in the $x-z$ plane based on the *triplet-search* approach [1] and then confirms the initial track by adding the $y$ information from the two $\pm 5°$ tilted SciFi layers. Since each layer contains an average number of 400 hits, the algorithm copes with $400^3$ hit combinations to make the tracks. A constraint imposing the track originates from the $(x=0, z=0)$ point reduces the number of *tracklets* to about 100. Removing the ones that have been used in the *long* track reconstruction, the initial tracklets for the `Downstream` algorithm can be around 20.

[1] A general approach to scan all possible 3 hits combinations in different detector layers.

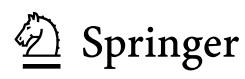

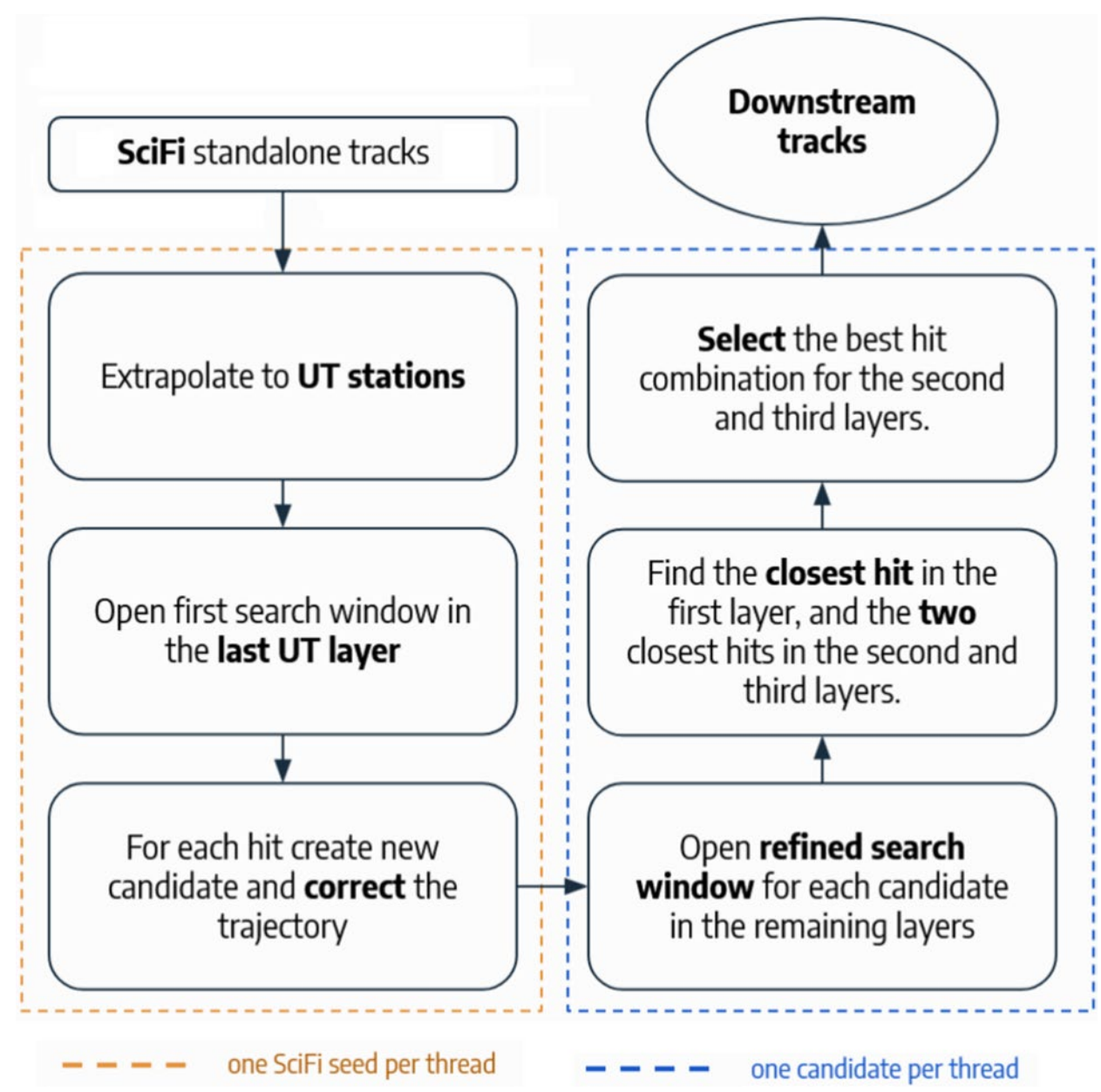

**Fig. 5** Flow chart of the `Downstream` algorithm (Dv0)

The change in the track direction by the magnetic field $\vec{B}$ is governed by the equation of motion

$$\frac{\mathrm{d}\vec{p}}{\mathrm{d}t} = q\vec{v} \times \vec{B}, \tag{1}$$

with $\vec{p}$, $q$ and $\vec{v}$ being the momentum, charge, and velocity of the particle. At LHCb, the magnetic field is predominantly in the $y$ direction, causing track deviations in the $x$–$z$ plane. This is modeled by a *kink* at a specific position along the $z$-axis—the *magnet point* $(x_{\text{magnet}}, y_{\text{magnet}}, z_{\text{magnet}})$. This point is parameterised using the coordinates $(x, y)$ and the track slopes $(t_x, t_y)$, where $t_x = px/pz = \mathrm{d}x/\mathrm{d}z$ and $t_y = py/pz = \mathrm{d}y/\mathrm{d}z$, of the seeds in the last SciFi station. An initial $q/p$ estimation also enters in the computation. Parameters are obtained by minimizing the difference between the predicted and true values of $z_{\text{magnet}}$ using simulated samples. The coordinates $x_{\text{magnet}}$ and $y_{\text{magnet}}$ are then calculated by straight-line extrapolation from SciFi, where a correction depending on the position and slope of the track is applied to correct the non-zero magnetic field in $x$ and $z$. In Fig. 6 the distribution of the bias vs the $z_{\text{magnet}}$ is shown from simulated events.

In a first version of the algorithm (Dv0) a search window[2] for compatible hits is opened in the last layer of the UT (*UTbX*) (see Fig. 2), this search window size is computed as a function of momentum, assuming the track originates from $(x = 0, z = 0)$. The parameters used to define the search window at each UT layer are derived from simulation studies.[3] For each hit in the search window of the last UT layer (*UTbX*), a new trajectory is computed using the hit position and $z_{\text{magnet}}$, and is then further extrapolated to the remaining UT layers (*UTaX*, *UTaU*, *UTbV*), opening new search windows whose sizes are defined as functions of momentum to account for multiple scattering effects. The closest hit in the first layer (*UTaX*) and the two closest hits in the second and third UT layers (*UTaU*, *UTbV*) are considered to form *downstream* candidates. Up to 10 closest hits are considered in the initial *UTbX* layer, and *tracklets* without any matched hits in this layer are skipped.

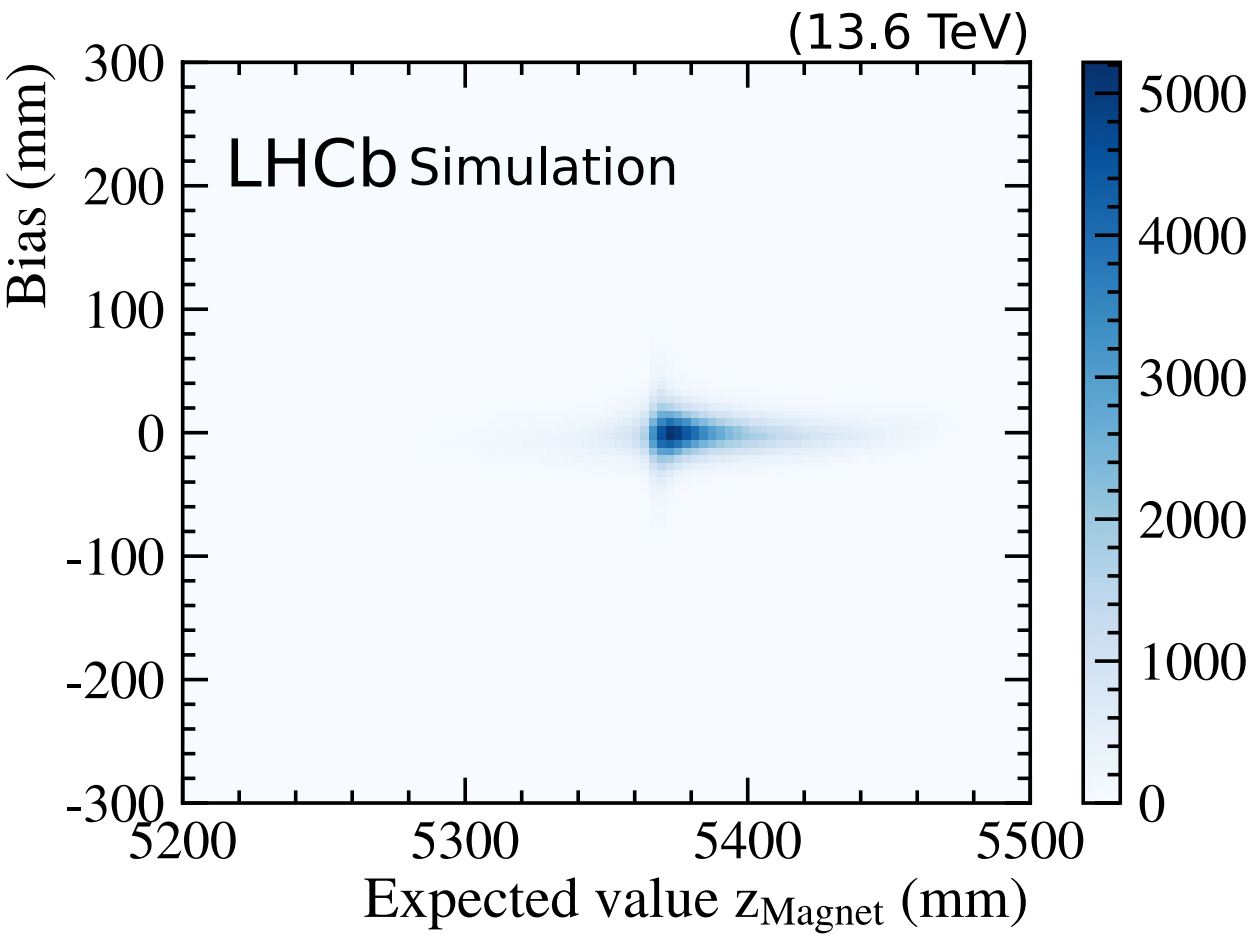

**Fig. 6** Distribution of the bias in $z_{\text{Magnet}}$

An improved version of the algorithm (Dv1) aiming for the reconstruction of new unknown particles with large masses, has also been developed. Assuming that downstream tracks point to the origin of the coordinate system in the definition of the first search window may have a negative impact on the tracking efficiency for decays, where tracks have very large opening angles. Therefore, an initial tracking step with a triplet search algorithm is applied to increase the efficiency of heavy particle decays which involve large opening angles. Once the magnet point is computed for each SciFi tracks, a parallelized triplet search is performed along with the hits in the *UTaX* and *UTbX* layers of the UT to create *downstream* candidates.

To optimize the throughput in this computationally expensive new configuration, UT hits are grouped into different *Y* sectors. Each SciFi track searches only for hits within compatible *Y* sectors, under the assumption that there is no magnetic field component in the *X* or *Z* directions. Once the triplet is defined in the axial layers of the UT, hits

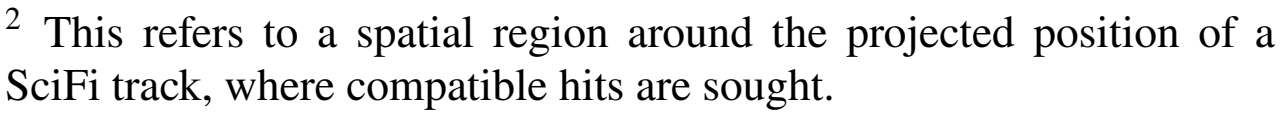
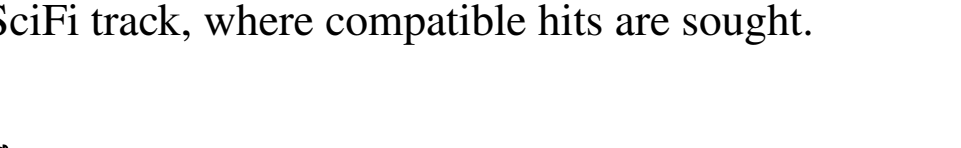

[2] This refers to a spatial region around the projected position of a SciFi track, where compatible hits are sought.

[3] These include the effects of the magnetic field and multiple scattering, and are used to parameterize the expected spread of hit positions.

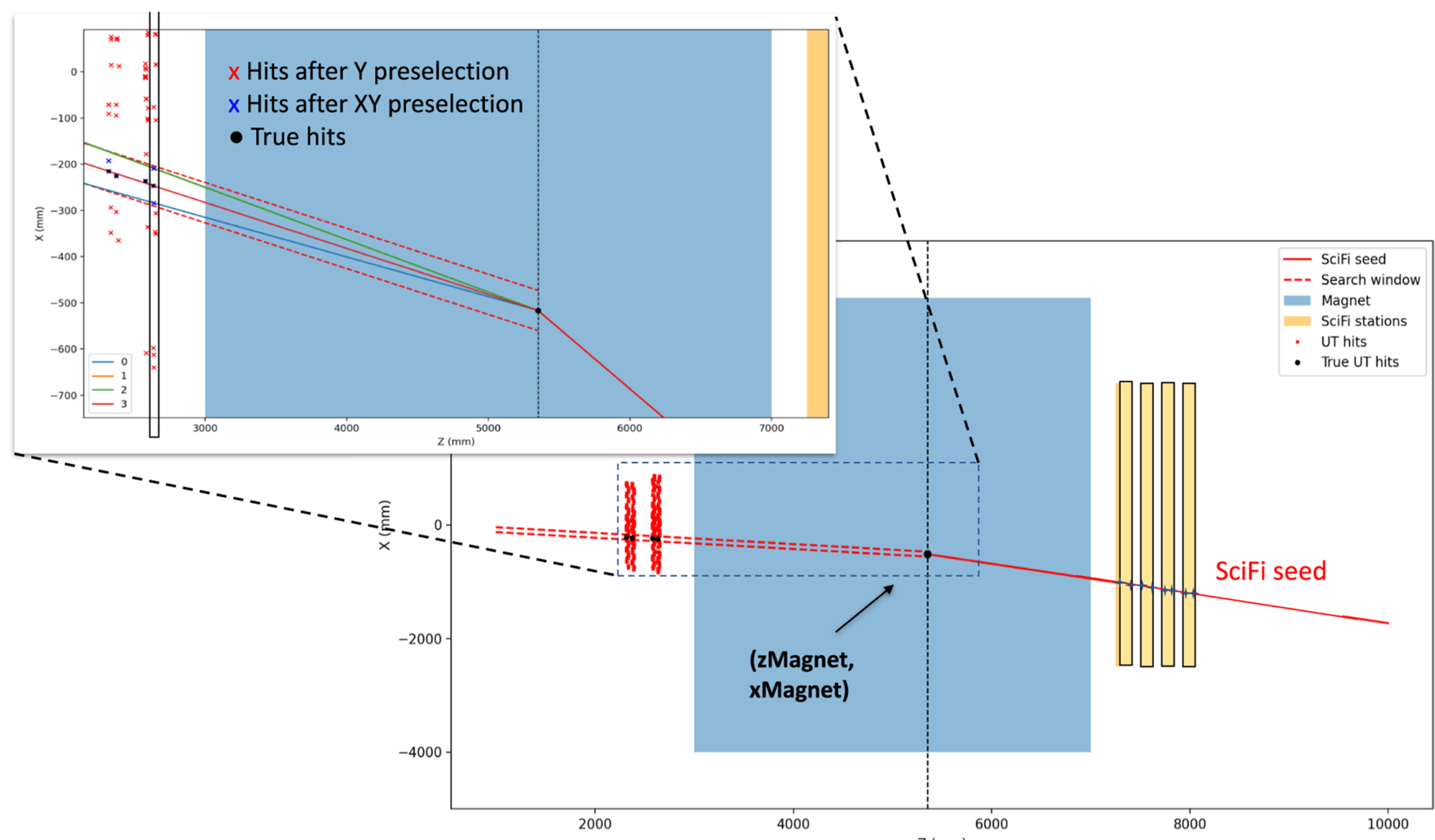

**Fig. 7** Strategy of the `Downstream` algorithm showing the used hits in the last UT layer and the Magnet Point ($x_{\text{Magnet}}$, $z_{\text{Magnet}}$) which is used to find the slope $t_x$. The slope is essentially the change in the $x$ position of the particle with respect to the change in the $z$ position from the Magnet Point to the UT stations

from the stereo layers (*UTaU* and *UTbV*) are added, and *downstream* tracks are constructed similar to the Dv0 of the algorithm,

The new version improves the tracking efficiency for decays of heavy BSM particles up to a factor of 4, and removes any mass dependence. It also enhances the efficiency for $\text{K}^0_\text{S}$ and $\Lambda$ by about 20% and 10%, respectively. It will be tested during the 2025 data-taking. Figure 15 shows the efficiency improvement for a new particle with a mass of 3 GeV/c$^2$ and 2 ns lifetime.

A final candidate is considered for each SciFi track based on the best combinations of hits according to a score computed with the distance between expected and real hit positions. This is achieved through a *clone killing* mechanism, which removes duplicate tracks that may have been created due to multiple hits associated with the same track. Iterating over all possible track pairs that have more than two shared UT hits or have the same SciFi segment, the candidates with lower scores are killed. This removes about 1% of the candidates.

One of the main features of the `Downstream` algorithm is the rejection of *ghost* tracks, which are fake tracks originated by spurious hits in the detector. This is explained in the following section.

Figure 7 shows an sketch of the algorithm strategy.

## Ghost-Killer Neural Network

Ghost tracks arise due to detector noise or reconstruction ambiguities. To suppress these tracks a streamlined feed-forward neural network (FFNN) with a unique hidden layer is used. The architecture includes as input six variables with information of the *downstream* track state ($x$,$y$,$t_x$,$t_y$,$q/p$,$\chi^2$), with $t_i$ the slope parameter (=d$i$/d$z$), and $\chi^2$ a quantity to evaluate the goodness of the track fit. Information of the SciFi track properties ($q/p$ and $\chi^2$) are also included, completing eight input variables. A hidden layer with a *ReLu* activation function [14] and 14 neurons is used. The optimal number of neurons in the hidden layer is obtained by studying the ghost rejection rate vs the throughput of the algorithm. A Sigmoid function [14] is used for the output to define a ghost probability score. Figure 8 shows the Ghost-killer architecture.

The training of the FFNN is performed using 10k $Bs \rightarrow \phi\phi$ simulated events, and the Adam optimisation algorithm [15] with the binary cross-entropy loss function. The loss is defined as:

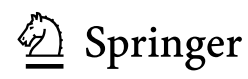

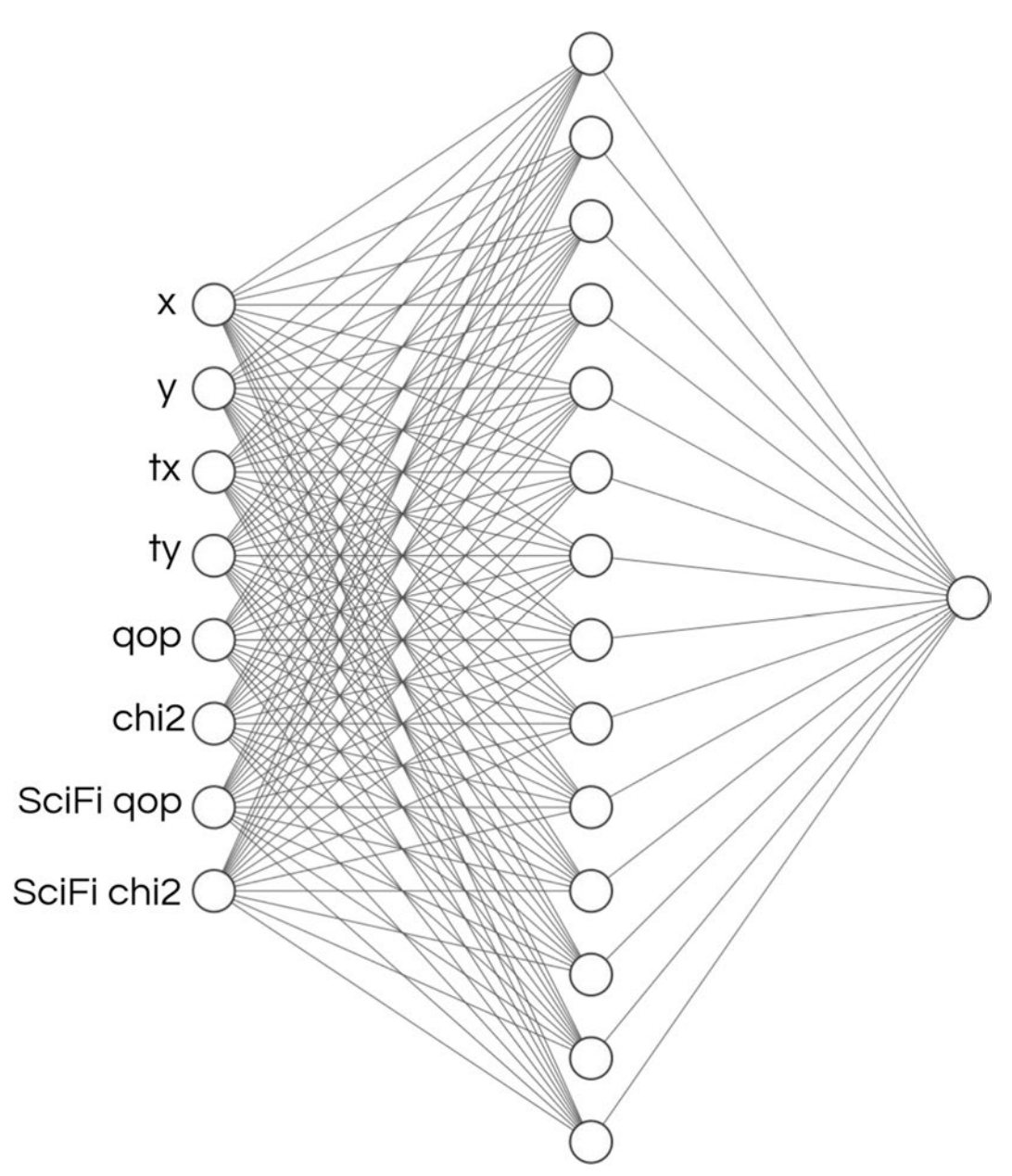

Fig. 8 Architecture of the FFNN used to remove ghost tracks

$$-\frac{1}{N}\sum_{i=1}^{N}\left[y_i \log(f(\mathbf{x}_i)) + (1-y_i)\log(1-f(\mathbf{x}_i))\right]$$

where $y_i$ is the true label of the $i$th example (1 for ghost, 0 for real track) and $f(\mathbf{x}_i)$ is the output of the NN for the $i$th example. The loss values in the final epoch ($\approx$ 18000) for both the test sample and train samples are found to be compatible. In Fig. 9 the classifier output for the Ghost-killer is shown.

During the training of the NN ghost killer, the clone-killing step of the tracking algorithm is disabled and the search window sizes are set to fixed, large values (100 mm for the first search window and 10 mm for the refined search window). This ensures that the training is not biased by the final track selection strategy and allows the search windows to be loosened later to accommodate differences between data and simulation without the need to retrain the NN.

## GPU Implementation

The algorithm is almost entirely executed on approximated 500 hundreds A5000 NVIDIA GPUs, which are hosted in up to 190 AMD EPYC dual-socket servers with 32 physical cores and 512 GB of RAM per socket. The CPUs are responsible for data copying and other supplementary services.

To take advantage of multi-threading in the GPU, data is processed in slices of events, each slice processing between 500 and 1000 events. Each CUDA thread block is mapped to a single event in the slice, and each thread within the block works on an independent candidate for reconstruction or selection.

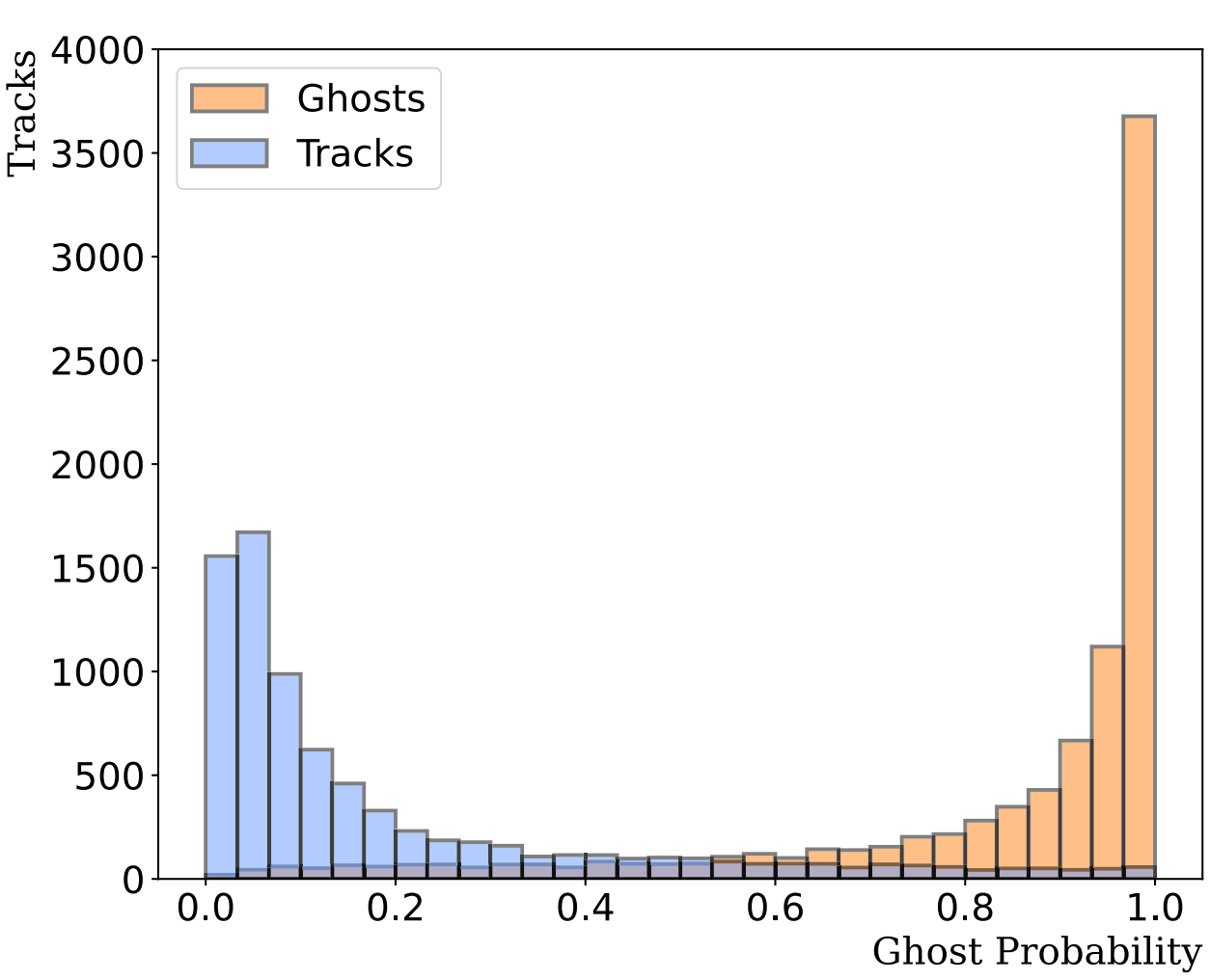

Fig. 9 Probability score given by the FFNN output

Threads within the same block can synchronize and share resources through shared memory. An RTX A5000 features 64 Streaming Multiprocessors (SMs), each containing 128 CUDA cores, for a total of 8192 CUDA cores. That means that 64 events can be processed in parallel.

Given the large hit combinatorics in the SciFi and UT detectors for making the track candidates, and the tight throughput constraints of the HLT1 sequence, the speed of the algorithm is mandatory.

To achieve this, static structures are used in the Ghost-killer NN architecture definition, fixing the number of input variables and the number of neurons. This allows the compiler to perform compile-time optimizations, such as utilizing registers instead of global memory, thereby improving the efficiency and performance of the implementation.

An additional technique, *loop unrolling*, which expands `for`-loops by replicating the loop body multiple times at compile time, is employed to eliminate data dependencies. This allows the compiler to further optimize the code and improve performance. It is achieved using the NVCC-specific `#pragma unroll` directive, which allows for partial or complete unrolling of loops. Fast mathematical functions from the CUDA library, such as `_fdividef` and `_expf`, are also utilized to accelerate floating-point operations. These functions are approximations, but they ensure the precision requirements of our use case. Furthermore, shared memory is leveraged to cache hits that are accessed multiple times during the combinatorics, reducing the overhead of repeated global memory accesses. In the following section, the performance of the algorithm is detailed.

## Algorithm Performance

The performance of track reconstruction of `Downstream` is evaluated using several figures of merit, in particular, throughput, momentum resolution, physics efficiency and ghost rejection. The throughput is evaluated, during the development phase, using a sample of simulated minimum bias proton–proton collisions which represent the collisions and detector response as in real data. The momentum resolution, physics efficiency and ghost rejection are evaluated using a bunch of dedicated simulated samples, which are representative of beauty, charm and strange, meson and baryon decays.

Results are validated using the data acquired by the LHCb during October 2024, with the UT running in global conditions and the `Downstream` algorithm included in the HLT1 sequence. Tracks of interest in physics events are expected to have a momentum above 5 GeV/c and a transverse momentum $p_{\mathrm{T}}$ above 0.5 GeV/c. These criteria are applied to determine the performance of `Downstream`.

An additional figure of merit correlated with the throughput is the power consumption of the algorithm. Sustainability is a subject which is taking increasing attention in view of the next generation of experiments and the need to analyze big amount of data. In this work we have evaluated the relative increase in power consumption of `Downstream`, to try to minimize it.

### Throughput

In Fig. 10, the performance of `Downstream` in terms of throughput is presented. It is related to the two reconstruction algorithms in the Allen sequence that determine *long* tracks: the `Matching` and `Forward` [16] algorithms. The minimal required throughput per GPU card for the LHCb HLT1 system is 60 kHz. The inclusion of `Downstream` in HLT1 slightly reduces the throughput by about 5 kHz for simulated and real data, when it is executed on a RTX A5000 card, as it can be seen in Fig. 10.

The throughput measurement in simulation is performed using the same method described in [11], but with an updated simulation sample. The throughput measurement on real data is done through a dedicated HLT1 testbench, which emulates the actual DAQ conditions and simulates the HLT1 processing as it occurs during real data-taking. During real data-taking, all detector readouts are packed and stored in a large buffer. HLT1 runs on top of this buffer and outputs to another buffer, from which HLT2 processes the data. In the testbench, dumped data files are used as input. They emulate both input and output services to feed data to HLT1 and collect its output. This setup allows to measure the most realistic throughput of HLT1 with this algorithm.

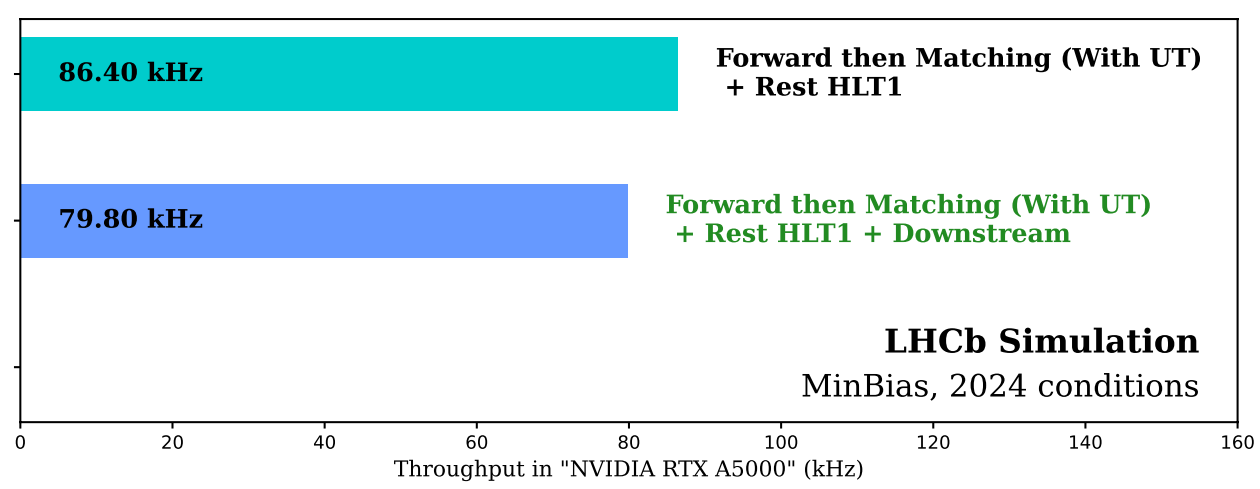

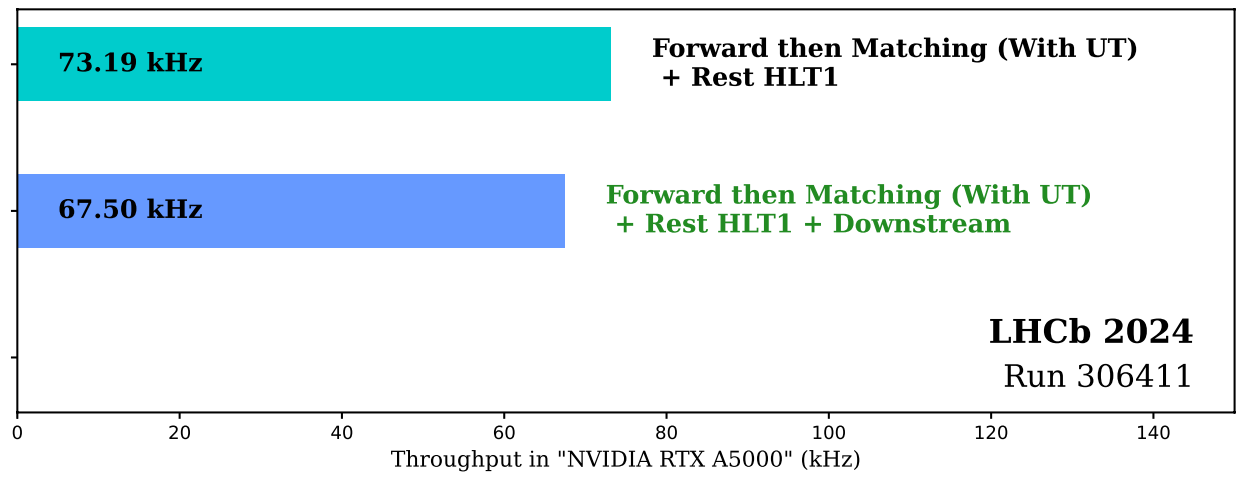

**Fig. 10** Throughput of the Allen sequence in simulation (top) and data (bottom) including the `Downstream` algorithm, executed on a RTA A5000 card and compared to other algorithms of the Allen sequence. The rest of the algorithms in the sequence refer to the remaining components of HLT1, which are in charge of the detector decoding, secondary vertex reconstruction, clustering, particle identification, etc.

### Momentum Resolution

The track momentum resolution is obtained using several simulated physics samples and is less than 2%. A slight increase is expected in real data due to the non-perfect simulation of the UT detector. Figure 11 shows the momentum resolution for the $\mathrm{B}^0 \rightarrow J/\psi K^0_s$ and $\Lambda_b \rightarrow \Lambda\gamma$ decay channels.

### Physics Efficiency

More than 75% of the *downstream* tracks are reconstructed by the algorithm in simulated samples. This has been verified for SM particles ($\Lambda$ and $\mathrm{K}^0_{\mathrm{S}}$) from different physics decay channels and from minimum bias events, and it is shown in Figs. 12 and 13.

Many new physics models predict long lived particles with masses larger than $\mathrm{K}^0_{\mathrm{S}}$ and $\Lambda$. During the October 2024 data-taking period, the `Downstream` algorithm was optimised for SM particles reconstruction, with tolerance windows in the UT tuned using simulated $\mathrm{K}^0_{\mathrm{S}}$ and $\Lambda$ trajectories to maximise the throughput. If the mass of the particle is much larger, the deflection due to the magnetic field is lower as shown in Fig. 14, and it can escape detection. To increase the algorithm's efficiency in case of unknown particles, the algorithm has been improved. The strategy has been modified to be able to search for hits in the first and last layers of the UT detector in parallel, without specifying any tolerance window. Special effort is done to keep similar throughput

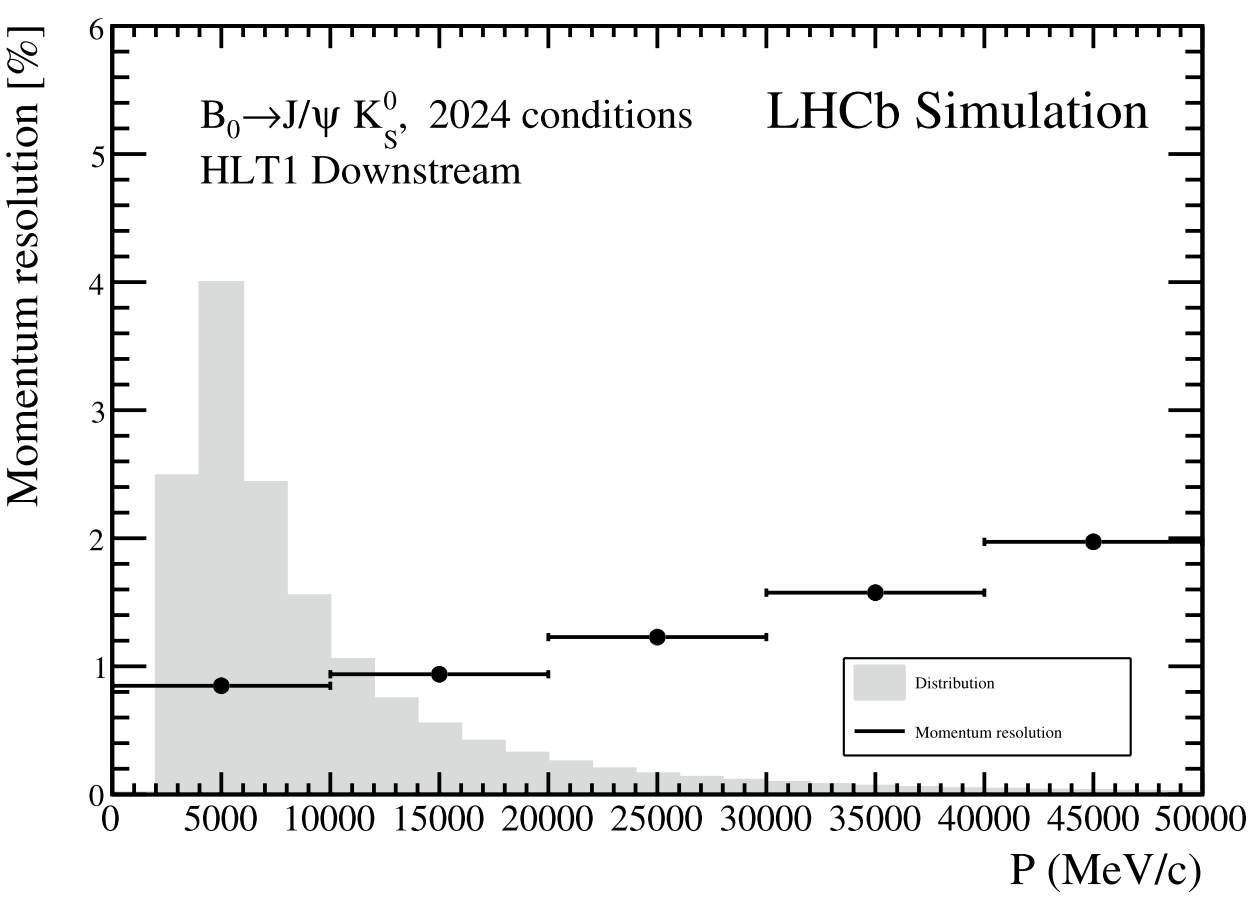

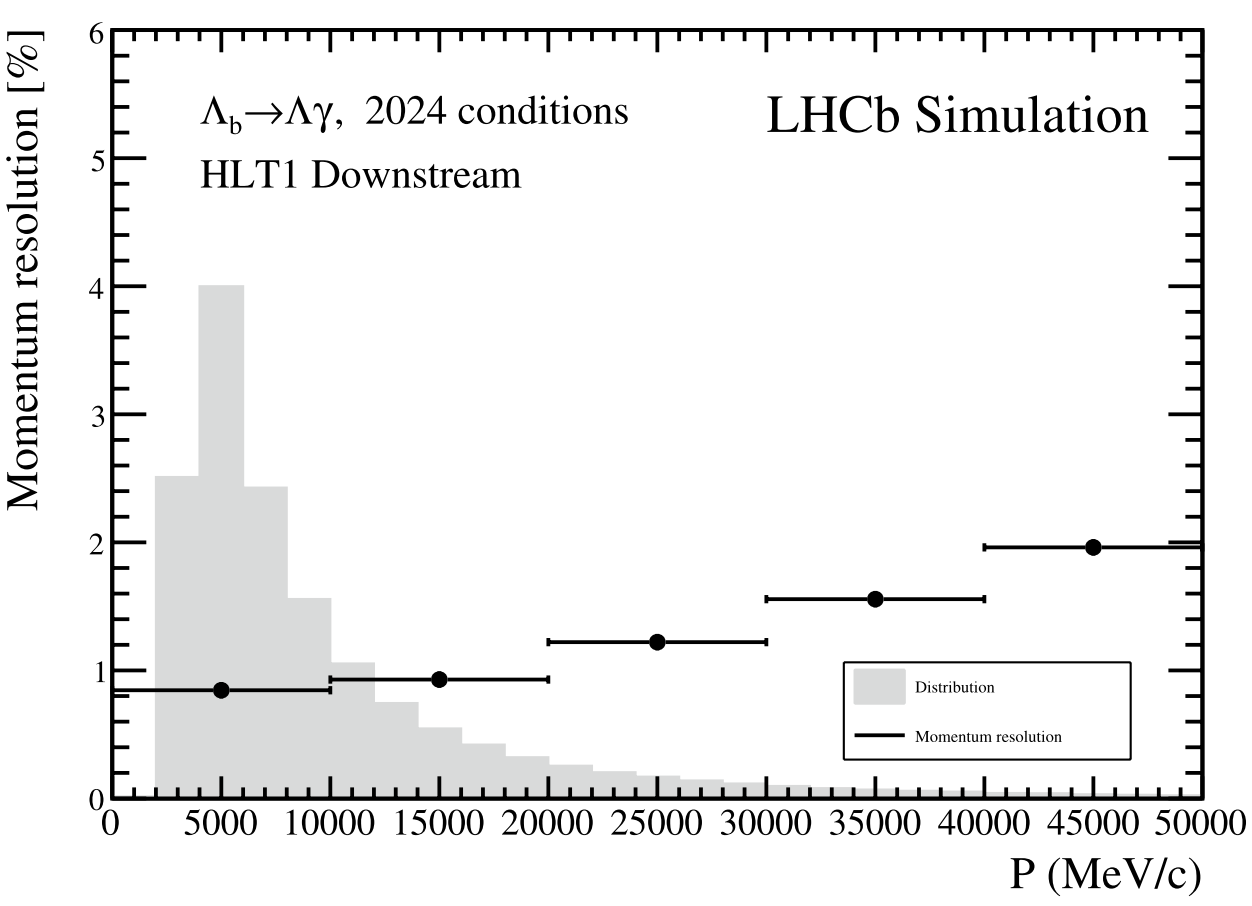

**Fig. 11** Momentum resolution of the tracks reconstructed by the `Downstream` algorithm using the simulated $\mathrm{B}^0 \rightarrow J/\psi K^0_S$ decay channel (top) and $\Lambda_b \rightarrow \Lambda\gamma$ decay channel (bottom). The gray distribution represents the normalized momentum distribution of all reconstructed tracks

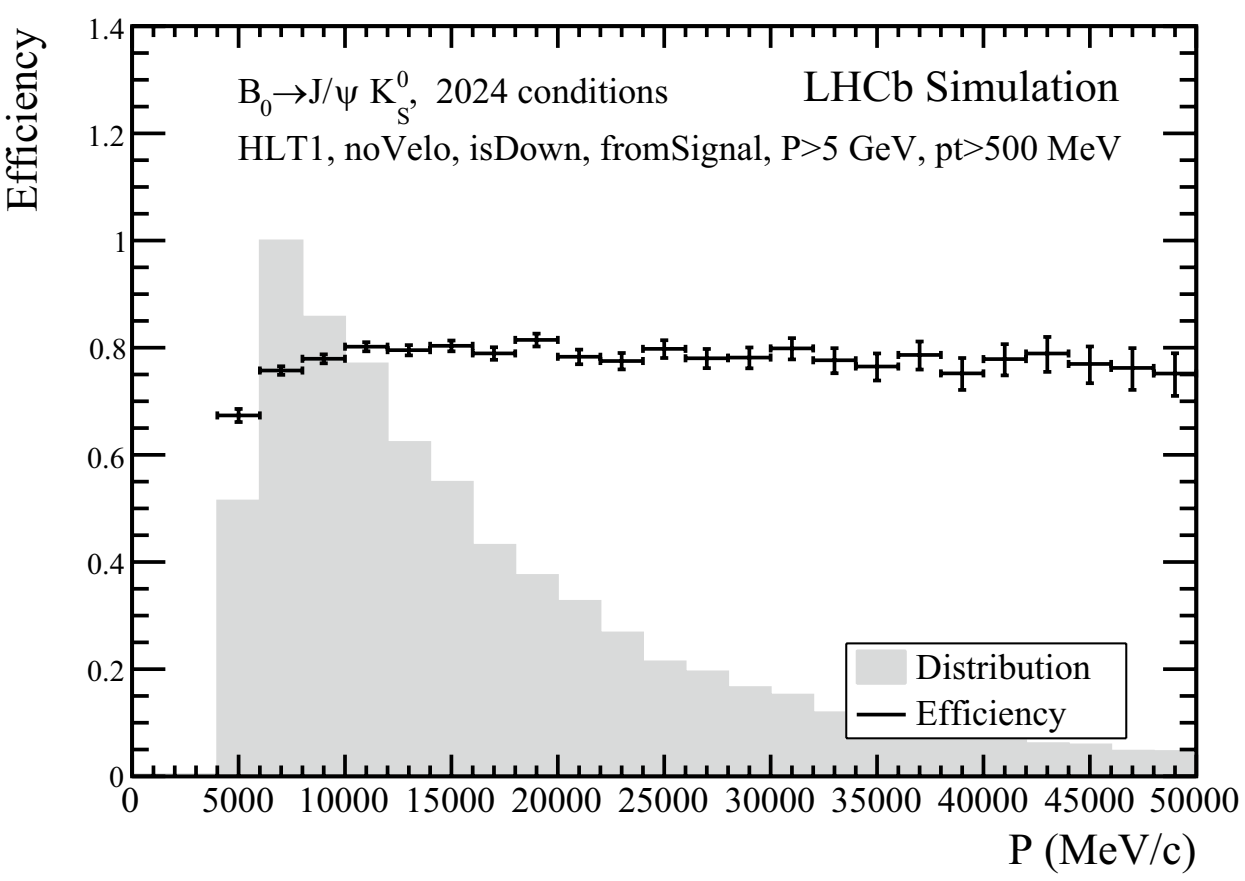

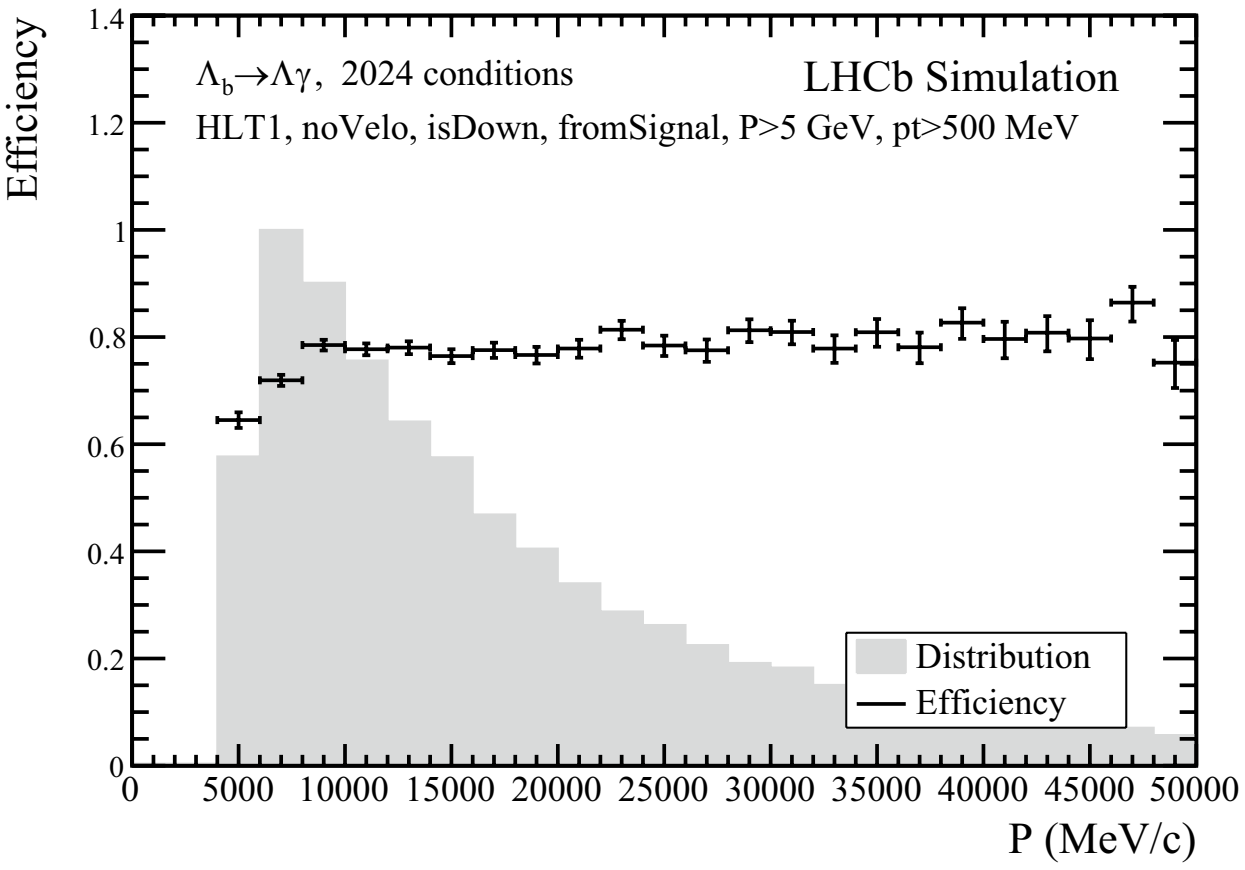

**Fig. 12** Efficiency of the `Downstream` algorithm as function of the momentum to reconstruct $\mathrm{K}^0_\mathrm{S}$ from $B^0$ decays (top) and $\Lambda$ from $\Lambda_b$ decays (bottom). Similar efficiencies are obtained for other decay channels

even if the combinatoric is larger, as explained in “Algorithm Design” section Algorithm Design.

The efficiency comparison for the nominal Dv0 version and improved Dv1 is shown in Fig. 15, for a new scalar dark bosons of 3 GeV/c$^2$ mass. As one may observe, Dv1 shows similar efficiency in heavy BSM decays as in $\mathrm{K}^0_\mathrm{S}$ and $\Lambda$ decays, overcoming the limitations of Dv0, which is optimized only for $\mathrm{K}^0_\mathrm{S}$ and $\Lambda$ decays. This brings more robustness to the tracking algorithms, removing the dependence with the particle mass. This new configuration will be used during the 2025 data taking period.

## Ghost Rejection

Distributions of the ghost rate vs the momentum are shown in Fig. 16 for two simulated decay channels. The ghost rate is defined as the fraction of reconstructed tracks that do not correspond to any real particle. The ghost-killer NN assigns a score to each reconstructed track, representing the probability that the track is a ghost. To suppress the ghost rate, a filter is applied that requires

$$\text{NN Response} < \text{Threshold}.$$

The threshold applied to the ghost-killer NN in these figures is 0.5. Lower threshold can give a ghost rate up to 40%, reducing the efficiency of the algorithm.

## Power Consumption

The effect on the power consumption from the execution of `Downstream` algorithm in the HLT1 sequence is studied in the following and shown in Fig. 17. Several techniques are employed to measure the power consumption including the use of a metered power distribution unit (PDU[4]) within the rack, analysis of device driver outputs (e.g.,

[4] An APC PDU AP8858EU3 is used in this work.

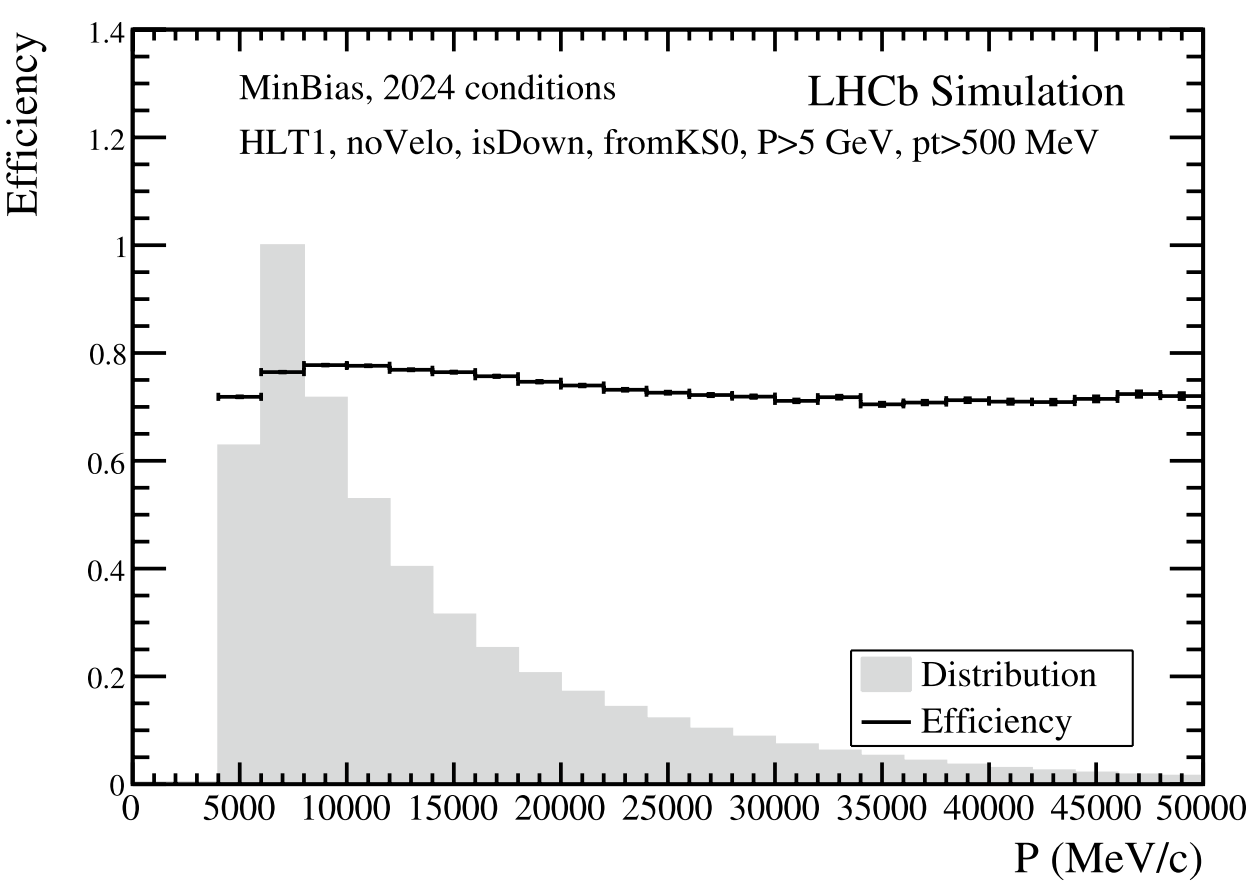

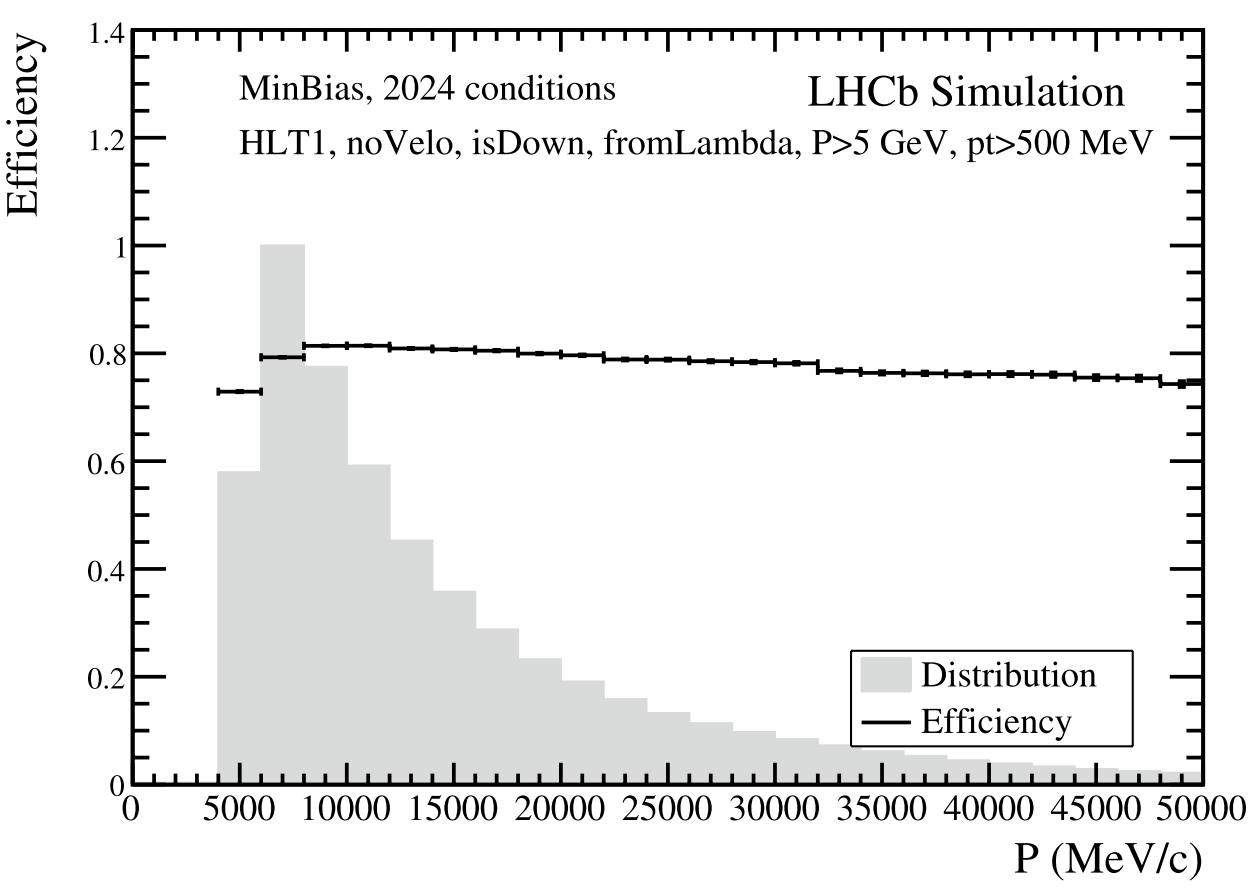

**Fig. 13** Efficiency of the `Downstream` algorithm as function of the momentum to reconstruct $K^0_S$ (top) and $\Lambda$ (bottom) from minimum bias events. The gray distribution represents the normalized momentum distribution of all reconstructed tracks from $K^0_S$ (top) and $\Lambda$ (bottom) decays in this sample

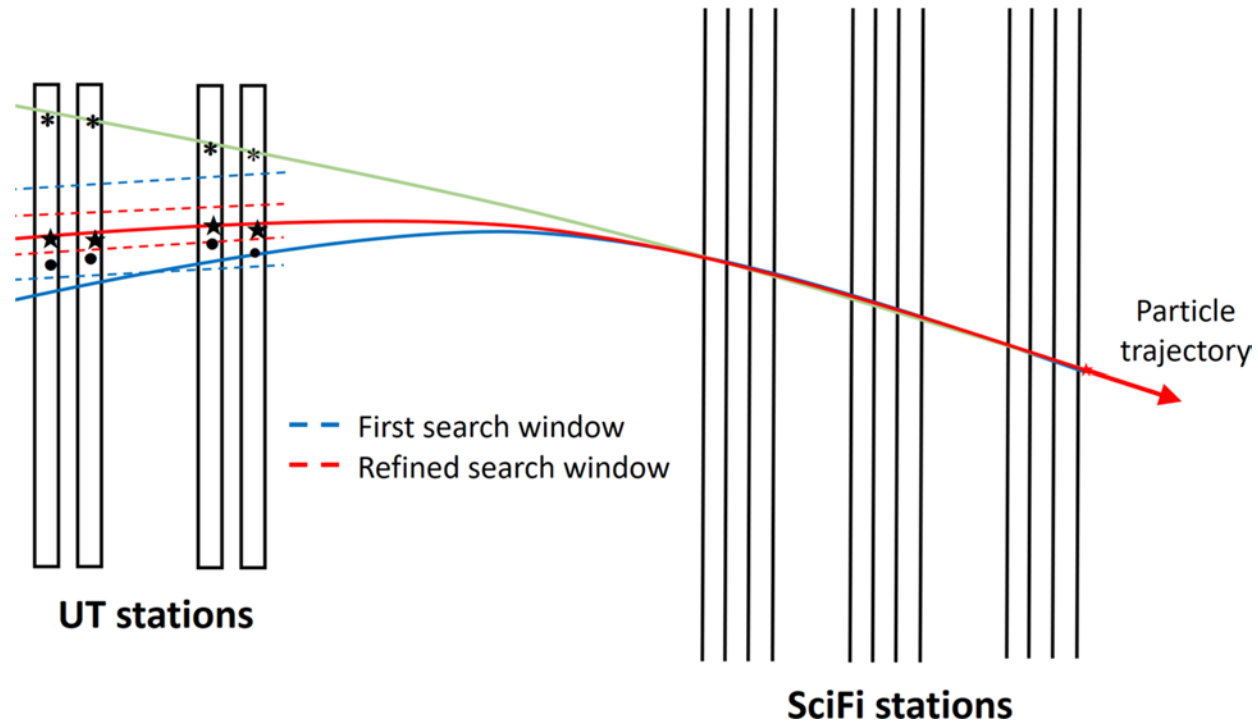

**Fig. 14** Effect of the search window in the Dv0 of the `Downstream` algorithm, tuned with $K^0_S$ and $\Lambda$ particles, on a new unknown heavier particle (represented in green, detector hits with asterisks). Two candidate trajectories for a SM particle are drawn in red and blue (bullet and star hits) according to the first and refined search window. The less pronounced curvature for a heavier BSM particle can make the search window to be non optimal. This inefficiency has been remedied in the Dv1 of the algorithm

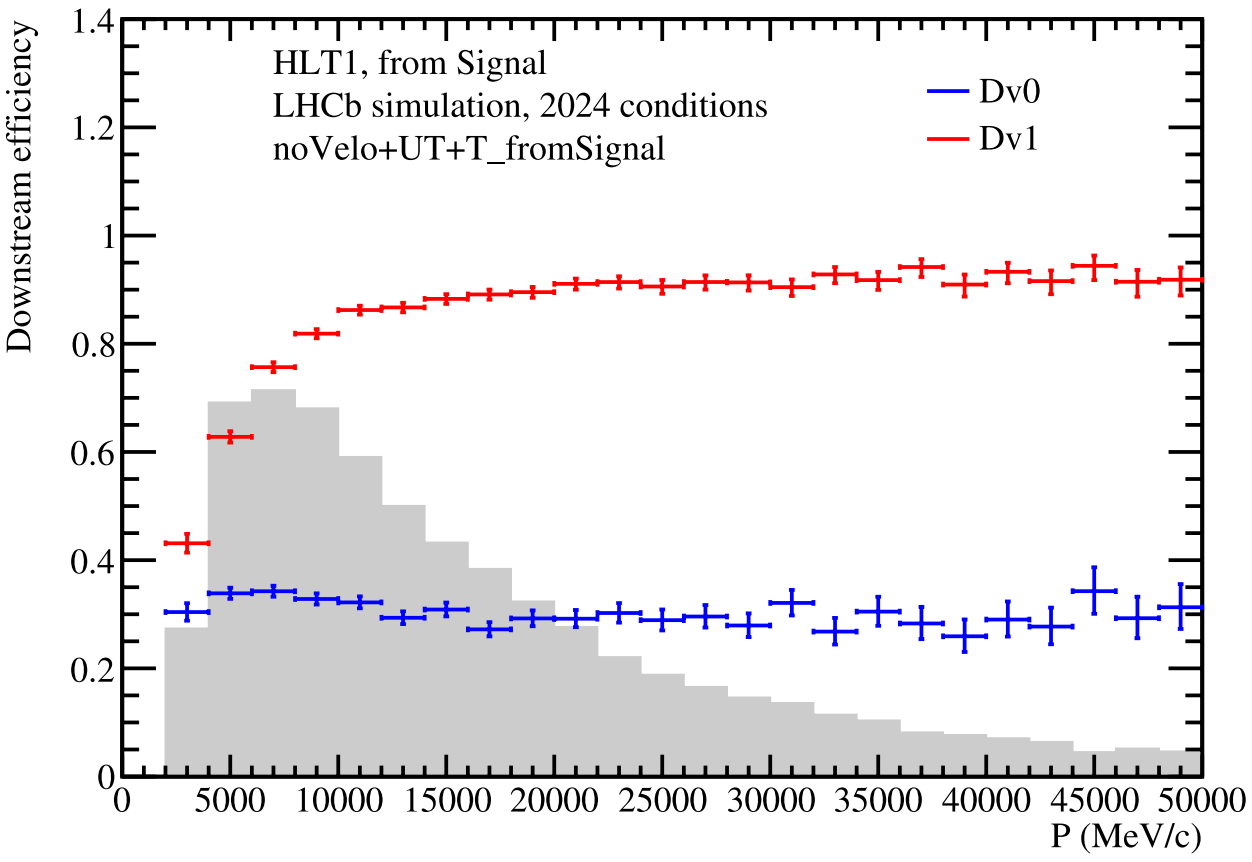

**Fig. 15** Efficiency comparison of the versions Dv0 and Dv1 of the `Downstream` algorithm for a new dark boson particle of mass 3 GeV/c$^2$ and lifetime of 2 ns. The gray distribution represents the normalized momentum distribution of all reconstructed signal tracks in the sample

Nvidia DCGM), monitoring CPU performance counters and extracting the estimated power consumption using ACPI system. Consistent results are obtained across all methods.

Only 6% increase in energy consumption is observed due to the highly optimized throughput of the `Downstream` sequence. The energy consumption of the system is correlated with the throughput,[5] as it can be inferred from Fig. 17.

## Vertexing Two *Downstream* Tracks

The vertexing of two *downstream* tracks consists of reconstructing the secondary vertices with two *downstream* tracks using a Newton–Raphson method and is explained in the following.

The nonzero magnetic field in the $y$ direction between the UT and the vertex position must be taken into account to avoid any bias in the $t_x$ slope of the track state at the origin vertex. In HLT1, the *downstream* track extrapolation cannot use a Runge–Kutta method due to computational costs. Instead, a second-order polynomial for $(z - z_0)$ is used to determine the $x(z)$ and $y(z)$ coordinates, where $z_0$ is the midpoint of the UT station. A parameter in the second-order term ($\gamma$) accounts for the magnetic field. This parameter is obtained by constraining the $t_x$ slope at the origin vertex (ovtx):

[5] The execution time of the system is inversely proportional to the throughput.

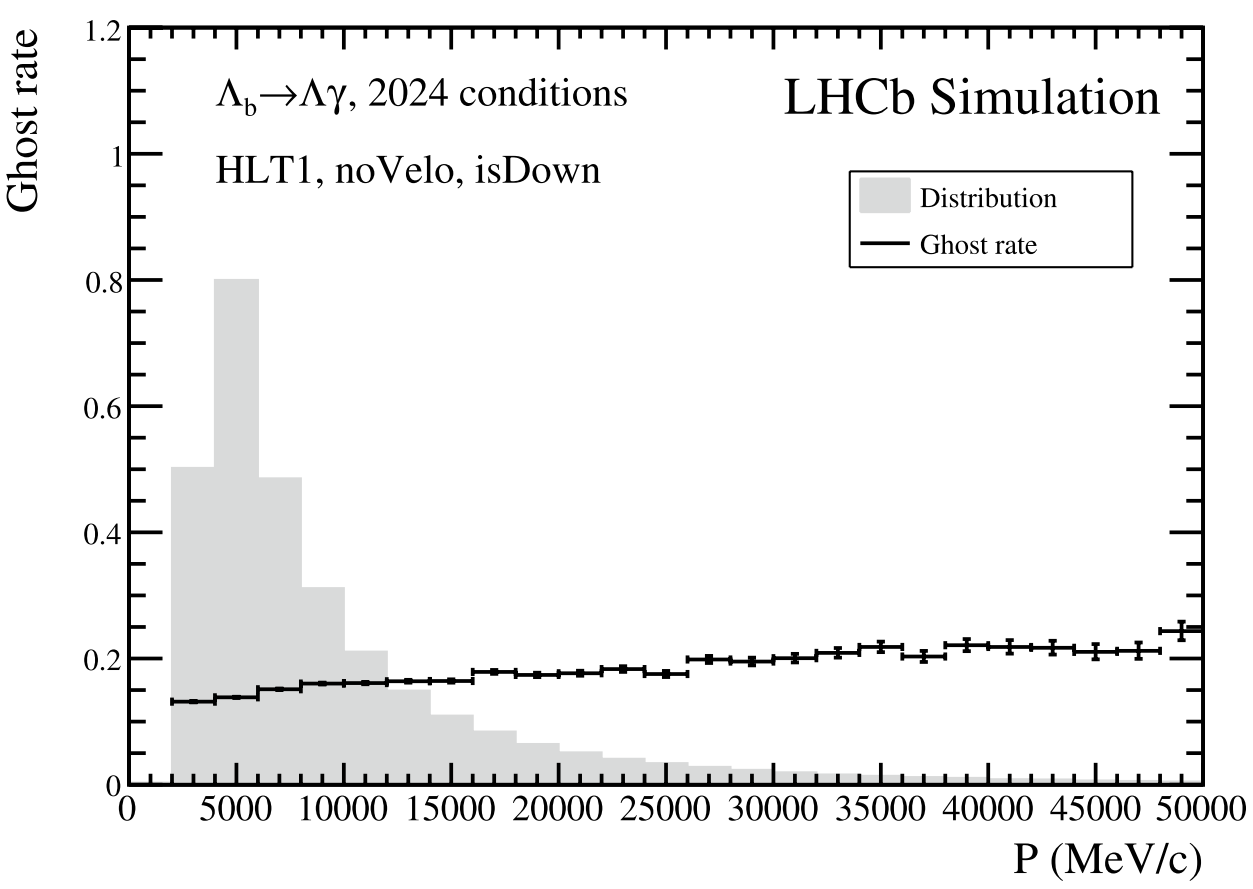

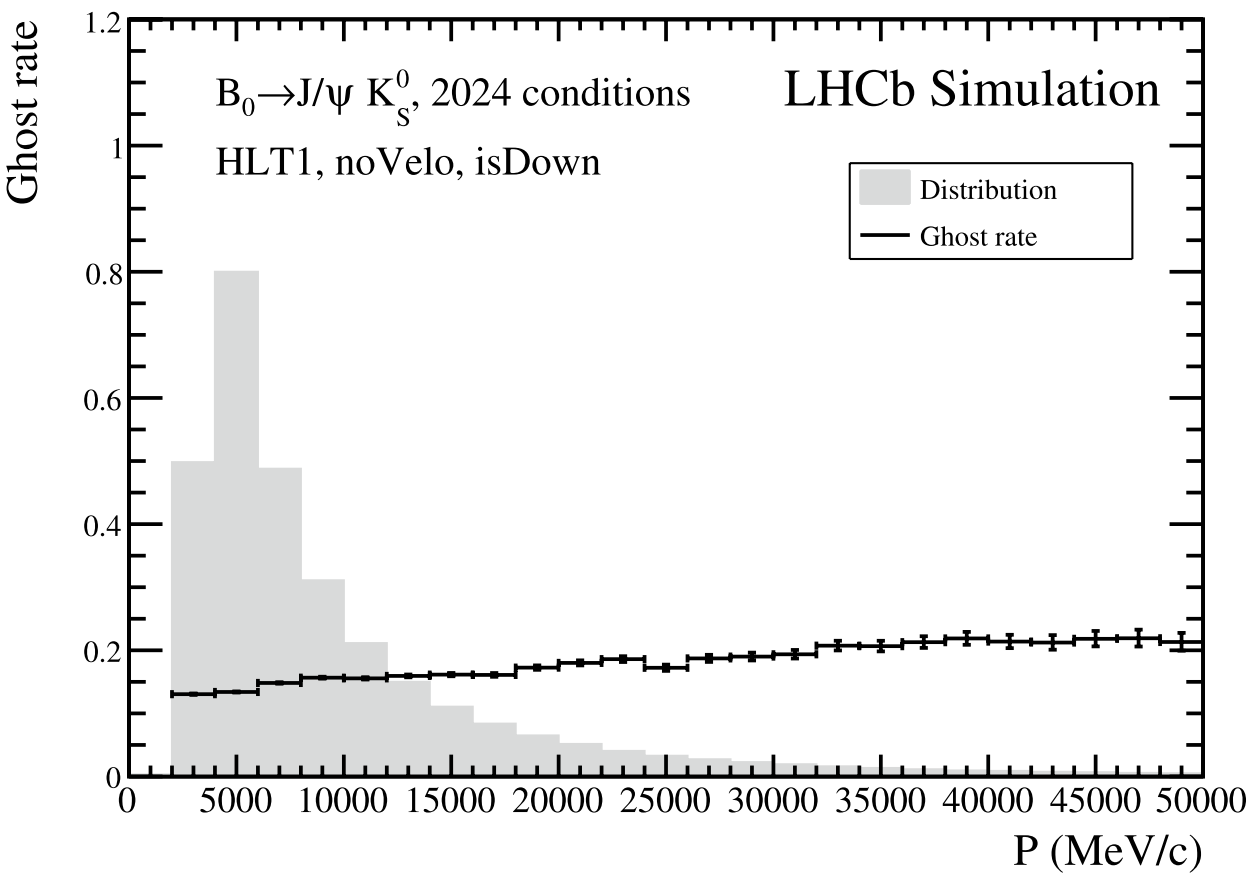

**Fig. 16** `Downstream` tracking ghost rate using $\Lambda_b \rightarrow \Lambda^0\gamma$ (top) and $B_0 \rightarrow J/\psi K^0_S$ (bottom) samples, as function of $p$. The gray distribution represents the normalized momentum distribution of all reconstructed tracks in the corresponding samples

$$t_x(\mathrm{ovtx}_z) = \left|\frac{\mathrm{d}x}{\mathrm{d}z}\right|_{z=\mathrm{ovtx}_z} = \mathrm{ovtx}_{t_x}. \tag{2}$$

Following simulation studies, the $\gamma$ parameter can be parametrized as function of $q/p$,

$$\gamma(q/p) = -\mathcal{MP}(-1.5 \times 10^{-8} + 9.3 \times 10^{-3} q/p), \tag{3}$$

where $\mathcal{MP}$ corresponds to the magnet polarity. This correction term is applied to the track extrapolation before determining the vertex. The trajectory of each track in state$_i$ given by $(x_i, y_i, z_{UT}, t_{xi}, t_{yi}, q/p_i)$ is thus

$$\mathrm{Traj}_i(z) = \begin{bmatrix} x_i + t_{xi}(z - z_{UT}) + \gamma_i(z - z_{UT})^2 \\ y_i + t_{yi}(z - z_{UT}) \\ z \end{bmatrix} \tag{4}$$

A Newton–Raphson method [17] is applied to find the point of closest approach between two extrapolated tracks, which is then used as the position of the secondary vertex. This method converges quickly, typically finding the solution in three iterations. The bias in the $z$ and $x$ positions of the vertex obtained using this technique are 14 cm and 1 cm, respectively. Figure 18 shows the improvement in the mass distributions for $K^0_S$ and $\Lambda$ particles when applying vertex reconstruction to simulated samples. The values are compared with the nominal (PDG) ones.

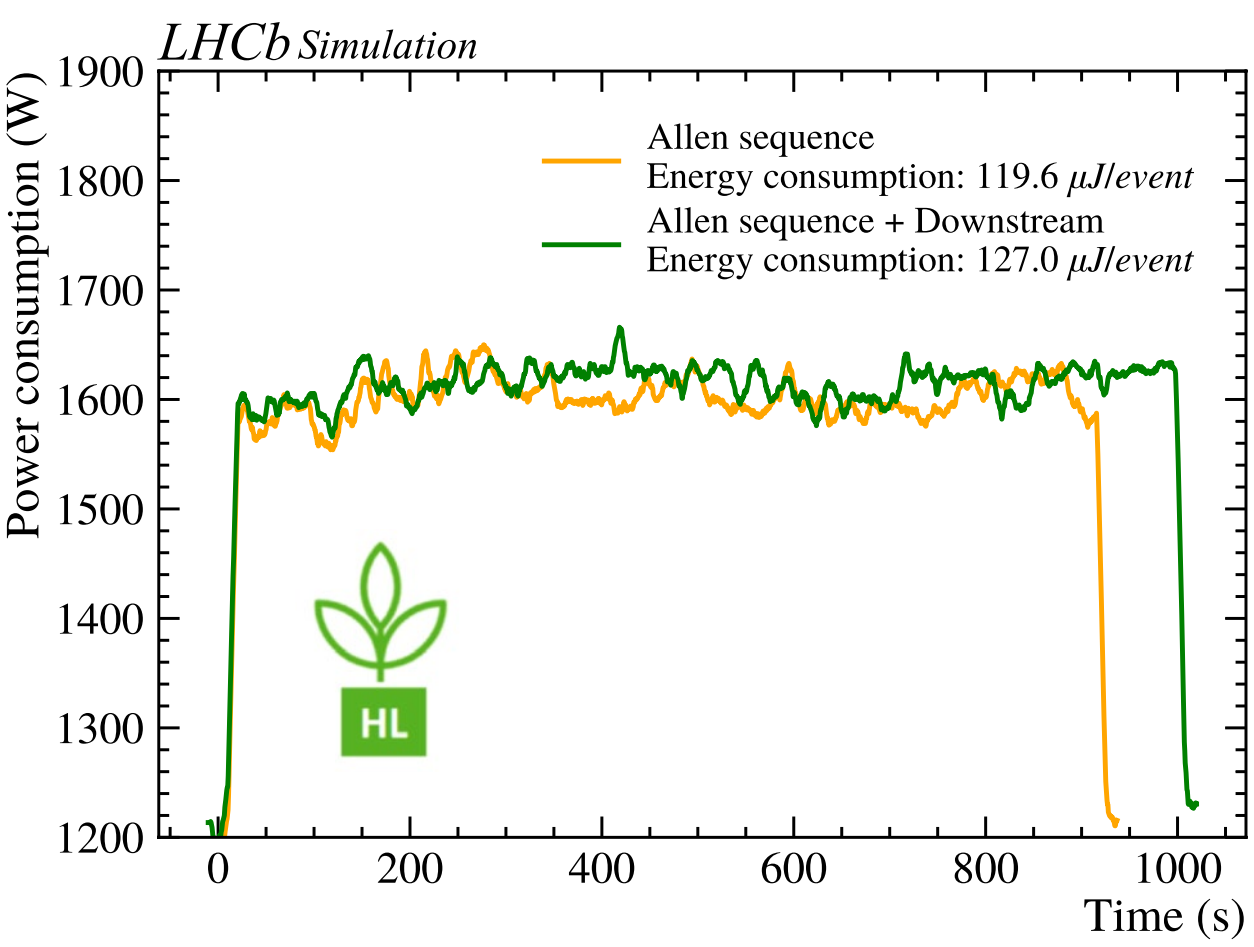

**Fig. 17** Power consumption with `Allen` software running over 3.2M $B_s \rightarrow \phi\phi$ events without (blue) and with (orange) `Downstream` algorithm. The power consumption is measured using metered rack PDU AP8858EU3 with an average readout frequency of 2 Hz. The moving average filter with window of 20 points is applied. The measurements are obtained using the NVIDIA RTX 6000 Ada Generation GPU card

## Expected Physics Impact

Using LHCb simulations with 2024 data-taking conditions, the effect of the `Downstream` and vertexing algorithms has been studied using several decay channels involving $K^0_S$ and $\Lambda$ decays. Two HLT1 trigger lines have been developed to select very detached downstream secondary vertices. The trigger lines rely on four different fast NNs with similar architectures to the one described in "Ghost-Killer Neural Network" section Ghost-Killer Neural Network one to select $K^0_S$ decays, one to select $\Lambda$ decays, one to distinguish detached $K^0_S$ decays from prompt $K^0_S$ decays, and one to distinguish detached $\Lambda$ decays from prompt $\Lambda$ decays. The inputs to these NNs include track parameters of the two secondary tracks, vertex parameters, and parent track parameters.

These NNs consist of a single hidden layer with 32 nodes and an output layer with a single node.[6] Secondary tracks are required to have opposite charges. A cocktail of simulated

[6] The increase in the throughput of the Allen sequence due to downstream vertexing and these lines is about 4%.

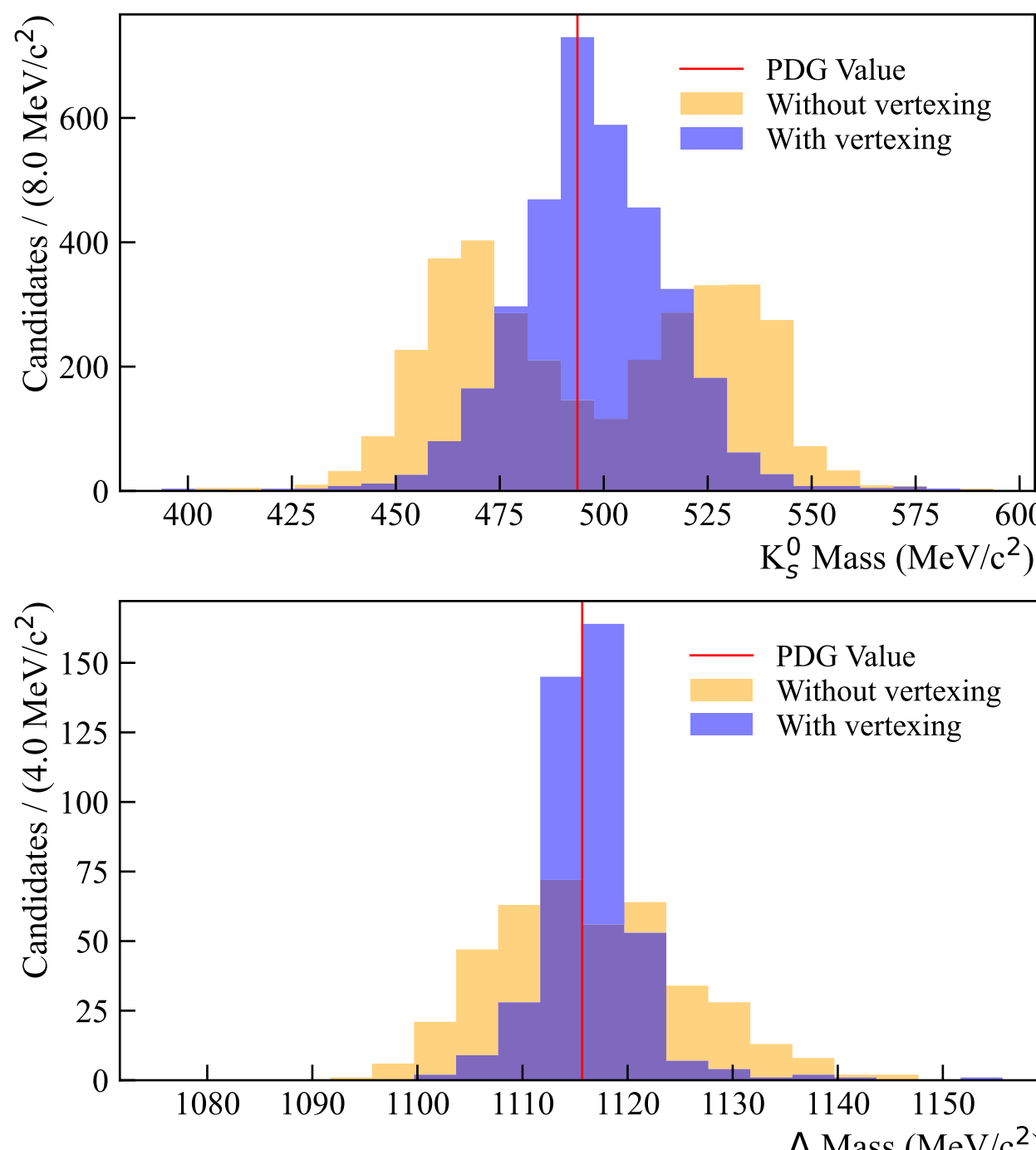

**Fig. 18** $K^0_S$ (top) and $\Lambda$ (bottom) mass distributions before and after applying the vertex reconstruction as explained in the text. Values are compared with nominal (PDG) ones from simulation. The mass resolution after applying the vertex reconstruction is 4 MeV/c$^2$ and 15 MeV/c$^2$ for $K^0_S$ and $\Lambda$ decays, respectively

physics channels, including detached $K^0_S$ and $\Lambda$ particles, is used for the training and testing of the trigger NNs. No overfitting is observed for the model. The NN output provides a classifier score, which determines the signal and background hypothesis of the *downstream* candidates.

Figure 19 shows the expected increase of efficiency when including triggered *downstream* tracks at the HLT1 level. The NN threshold for background suppression has been set to 0.5. Some of these channels, such as charm decays decaying into two $K^0_S$, or $b$ and charm baryon decays, are only available for *downstream* tracks, and the efficiency increase is 100%.

## Conclusion

The UT detector has been successfully installed and commissioned during the Summer 2024, providing proper *long*-tracks reconstruction during the Autumn run. During 2 weeks in October 2024, and following the installation of 163 additional GPUs, the newly developed `Downstream` algorithm was included in the full HLT1 sequence.

This algorithm, implemented within the Allen framework at the first level of the LHCb trigger, demonstrates the ability to reconstruct and select highly displaced vertices in real time at a 30 MHz data rate. It achieves high efficiency and low ghost track rates, with significant throughput facilitated by a fast and optimized neural network. Furthermore, its contribution to the overall power consumption of the full Allen sequence has been quantified.

**Fig. 19** Expected efficiency increase for different triggered $K^0_S$ and $\Lambda$ channels due to the inclusion of the `Downstream` algorithm. The blue band represents the *long*-track candidates, while the green one adds the selections via *downstream* tracks

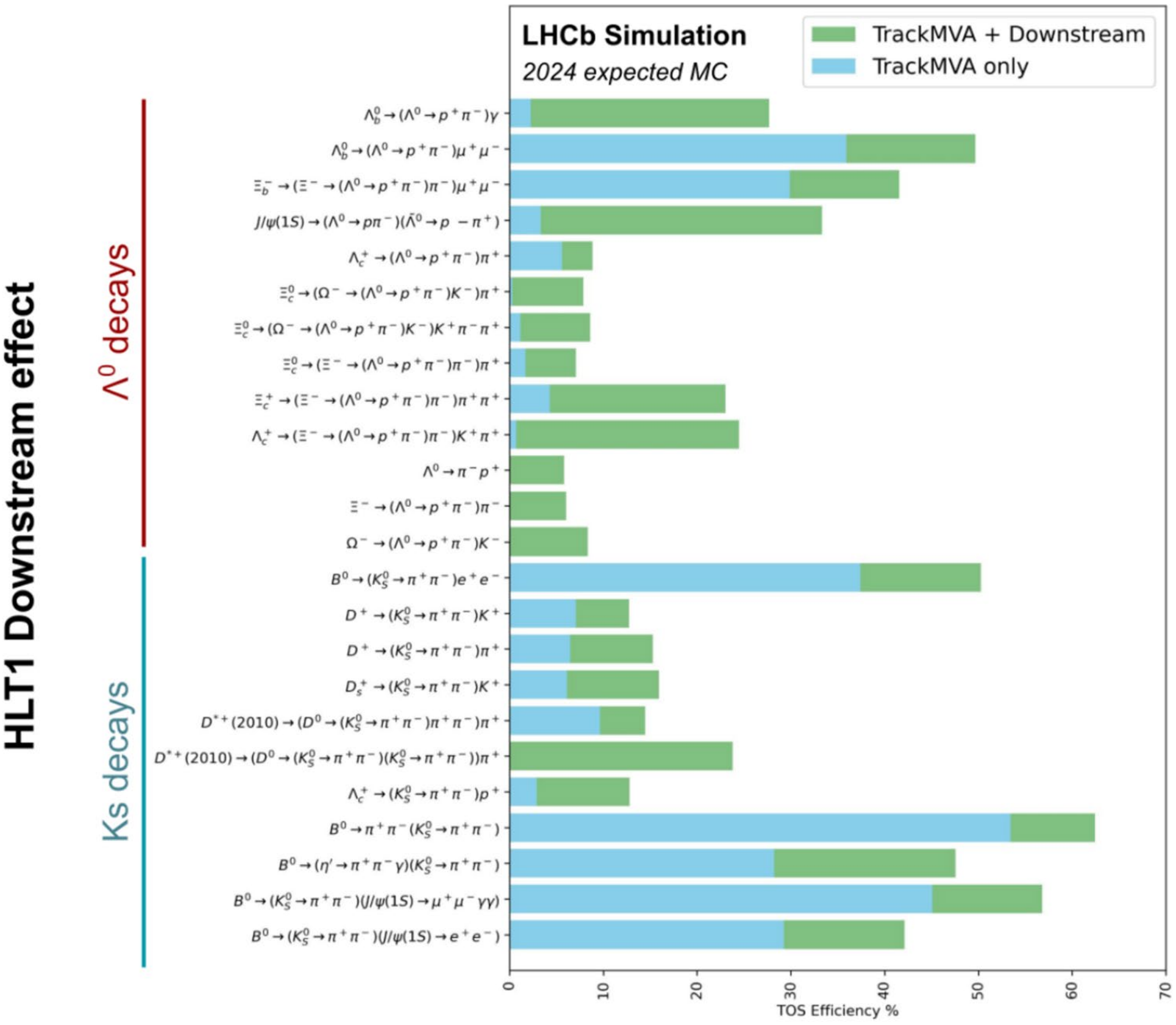

The algorithm's performance was validated using real data collected in October 2024, successfully reconstructing $\Lambda$ and $\mathrm{K}^0_\mathrm{S}$ particles that do not interact with the VELO detector. This capability is expected to have a high impact on LHCb physics during the 2025 data-taking period. Moreover, a new version of the algorithm extends its applicability to the reconstruction of heavier long-lived particles beyond the Standard Model, opening an unexplored phase-space and offering exciting opportunities for future discoveries [18].

**Acknowledgements** We thank LHCb's Real-Time Analysis project for its support, for many useful discussions, and for reviewing an early draft of this manuscript. We also thank the LHCb computing and simulation teams for producing the simulated LHCb samples used to benchmark the performance of the algorithm presented in this paper. The development and maintenance of LHCb's nightly testing and benchmarking infrastructure which our work relied on is a collaborative effort and we are grateful to all LHCb colleagues who contribute to it. We acknowledge the support from the Spanish Ministry of Science and Innovation, in particular via the project TED2021-130852B-I00, from TIFR (Mumbai) and from CONEXION AIHUB-CSIC, the US NSF cooperative agreement OAC-1836650 (IRIS-HEP) and the Simons Foundation.

**Author Contributions** Jiahui Zhuo and Brij Kishor Jahal are the main developers of the downstream tracking and vertexing algorithms. Valerii Kholoimov has provided the simulation samples, and performed the throughput tests. He created most of the figures in the document. Volodymyr Svintozelskyi has performed the power consumption measurements and has contributed to the tracking and vertexing algorithms development. Arantza Oyanguren is the team leader of this work who created the project, and the main editor of the document.

**Funding** Open Access funding provided thanks to the CRUE-CSIC agreement with Springer Nature.

**Data Availability** No datasets were generated or analysed during the current study.

## Declarations

**Competing interests** The authors declare no competing interests.

## References

1. The LHCb Collaboration, Batpista J et al (2020) LHCb upgrade GPU high level trigger technical design report. Tech. rep., CERN, Geneva. https://cds.cern.ch/record/2717938. Accessed 23 Apr 2025
2. The LHCb Collaboration, Aaij R (2024) The LHCb Upgrade I. J Instrum 19(05):P05065. https://doi.org/10.1088/1748-0221/19/05/P05065
3. The LHCb Collaboration, Alves AA Jr (2008) The LHCb Detector at the LHC. J Instrum 3(08):S08005. https://doi.org/10.1088/1748-0221/3/08/S08005
4. LHCb Collaboration, Aaij R (2025) Long-lived particle reconstruction downstream of the lhcb magnet. Eur Phys J C. https://doi.org/10.1140/epjc/s10052-024-13686-6
5. Svintozelskyi V (2024) Faraway algorithm to reconstruct and trigger vertices from long-living particles at LHCb. https://cds.cern.ch/record/2914251. Accessed 23 Apr 2025
6. Christian Bierlich S et al (2022) Codebase release 8.3 for PYTHIA. SciPost Phys Codebases. 8–r8.3. https://doi.org/10.21468/SciPostPhysCodeb.8-r8.3. https://scipost.org/SciPostPhysCodeb.8
7. Lange DJ (2001) The Evtgen particle decay simulation package. Nucl Instrum Meth A 462(1):152–155. https://doi.org/10.1016/S0168-9002(01)00089-4
8. GEANT4 Collaboration, Agostinelli S et al (2003) Geant4—a simulation toolkit. Nucl Instrum Methods Phys Res A 506(3):250–303. https://doi.org/10.1016/S0168-9002(03)01368-8
9. Li P, Rodrigues E, Stahl S (2021) Tracking Definitions and Conventions for Run 3 and Beyond. Tech. rep., CERN, Geneva. https://cds.cern.ch/record/2752971. Accessed 23 Apr 2025
10. The LHCb Collaboration (2014) LHCb trigger and online upgrade technical design report. Tech. rep. https://cds.cern.ch/record/1701361. Accessed 23 Apr 2025
11. Aaij R et al (2020) Allen: a high-level trigger on GPUs for LHCb. Comput Softw Big Sci 4(1):7. https://doi.org/10.1007/s41781-020-00039-7
12. Jashal BK (2023) Triggering new discoveries: development of advanced HLT1 algorithms for detection of long-lived particles at LHCb. https://cds.cern.ch/record/2881886. Accessed 23 Apr 2025
13. Aiola S et al (2021) Hybrid seeding: a standalone track reconstruction algorithm for scintillating fibre tracker at LHCb. Comput Phys Commun 260:107713. https://doi.org/10.1016/j.cpc.2020.107713
14. Lederer J (2021) Activation functions in artificial neural networks: a systematic overview. https://arxiv.org/abs/2101.09957. Accessed 23 Apr 2025
15. Kingma DP, Ba J (2017) Adam: a method for stochastic optimization. https://arxiv.org/abs/1412.6980. Accessed 23 Apr 2025
16. Bailly-Reyre A, Bian L, Billoir P, Pérez DHC, Gligorov VV, Pisani F, Quagliani R, Scarabotto A, Bruch DV (2024) Looking forward: a high-throughput track following algorithm for parallel architectures. IEEE Access 12:114198–114211. https://doi.org/10.1109/ACCESS.2024.3442573
17. Dedieu JP (2015) Newton-Raphson Method. In: Engquist B (ed) Encyclopedia of applied and computational mathematics. Springer, Berlin, pp 1023–1028
18. Gorkavenko V, Jashal BK, Kholoimov V, Kyselov Y, Mendoza D, Ovchynnikov M, Oyanguren A, Svintozelskyi V, Zhuo J (2024) LHCb potential to discover long-lived new physics particles with lifetimes above 100 ps. Eur Phys J C. https://doi.org/10.1140/epjc/s10052-024-12906-3